\documentclass[12pt]{article}
\usepackage[cp1250]{inputenc}
\usepackage[verbose,a4paper,tmargin=0.5in,lmargin=1in,rmargin=1in]{geometry}
\usepackage[pdftex]{graphicx}
\usepackage{amsfonts}
\usepackage{indentfirst}
\usepackage{latexsym}
\usepackage{amsmath}
\usepackage{amssymb}
\usepackage{psfrag}
\usepackage{eufrak}
\usepackage[mathscr]{eucal} 

\title{Null Killing vectors and geometry of null strings in Einstein spaces.}

\author{$\textrm{Adam Chudecki}^{*}$}

\begin{document}

\maketitle

$*$ Center of Mathematics and Physics, Technical University of Łódź           
\newline
Al. Politechniki 11, 90-924 Łódź, Poland, adam.chudecki@p.lodz.pl
\newline
\newline
\textbf{Abstract}. 

Einstein complex spacetimes admitting null Killing or null homothetic Killing vectors are studied. These vectors define totally null and geodesic 2-surfaces called the null strings or twistor surfaces. Geometric properties of these null strings are discussed. It is shown, that spaces considered are hyperheavenly spaces ($\mathcal{HH}$-spaces) or, if one of the parts of the Weyl tensor vanishes, heavenly spaces ($\mathcal{H}$-spaces). The explicit complex metrics admitting null Killing vectors are found. Some Lorentzian and ultrahyperbolic slices of these metrics are discussed.
\newline
\newline
\textbf{PACS numbers:} 04.20.Cv, 04.20.Jb, 04.20.Gz

\section{Introduction}

Complex methods in general theory of relativity have attracted a great interest for many years. Null tetrad formalism, twistor analysis and finally heavenly and hyperheavenly spaces ($\mathcal{H}$-spaces and $\mathcal{HH}$-spaces) play an important role in physics and mathematics. Presented work uses the heavenly and hyperheavenly formalism in analysis of the metrics admitting the null Killing vector. 

Hyperheavenly spaces ($\mathcal{HH}$-spaces) was introduced in 1976 in famous work \cite{biblio_1} by J.F. Plebański and I. Robinson as a natural generalization of the heavenly spaces ($\mathcal{H}$-spaces). Hyperheavenly spaces with cosmological constant $\Lambda$ are complex spacetimes with algebraically degenerate self-dual or anti-self-dual part of the Weyl tensor satisfying the vacuum Einstein equations with cosmological constant. The transparent adventage of hyperheavenly spaces theory is the reduction of Einstein equations to one, nonlinear differential equation of the second order, i.e. \textsl{hyperheavenly equation}. It seemed, that finding new real vacuum solutions of Einstein field equations with the signature $(+++-)$ was only a matter of time. It was enough to solve the hyperheavenly equations and then to find Lorentzian slices of respective complex spacetimes. This reaserch programme has been named \textsl{Plebański programme}. Unfortunatelly, obtaining the real slices appeared to be more difficult then anyone has ever suspected. In order to understand better the problem, the structure of hyperheavenly spaces together with their spinorial description has been investigated by J.F. Plebański, J.D. Finley III and co-workers \cite{biblio_4} - \cite{biblio_3}. Believing that symmetry of the spacetime simplifies the problem, Killing symmetries in heavenly and hyperheavenly spaces have been considered \cite{biblio_17} - \cite{biblio_dod_2}. Lorentzian slices remained elusive, except some examples \cite{biblio_3}, \cite{biblio_dod_16} - \cite{biblio_dod_33} and discussions \cite{biblio_5} no general techniques have been presented. Probably it was the reason why hyperheavenly machinery became less popular in nineties. 

Within five last years hyperheavenly spaces found their place in deep mathematical considerations. Their connection to Walker and Osserman geometry has been noticed in 2008. A few transparent results have been obtained with help of hyperheavenly formalism \cite{biblio_9, biblio_10}. It appeared, that $\mathcal{HH}$-spaces are the most natural tool in investigating real space of the neutral (ultrahyperbolic) signature $(++--)$. Moreover, a few works devoted to Killing symmetries in heavenly and hyperheavenly spaces appeared \cite{biblio_15} - \cite{biblio_dod_10}. These papers generalized the previous ideas of J.F. Plebański, J.D. Finley III, S. Hacyan and S.A. Sonnleitner. Between Killing vectors especially useful are these ones, which are tangent to self-dual null string. The existence of such (null) Killing vectors simplyfies the hyperheavenly equation, making it solvable in majority of cases.

The main aim of our work is to find all complex hyperheavenly and heavenly metrics admitting null homothetic and isometric Killing symmetry. Such metrics appear to be important in $(++--)$ real geometries. However, the existence of a null Killing vector appeared to be helpfull for finding the Lorentzian slices \cite{biblio_dod_3}. We want to develop this idea and examine all possible Lorentzian slices of the complex spacetimes admitting null Killing vector.

It is well known \cite{biblio_5} that if a complex spacetime admits any real Lorentzian slice then both self-dual and anti-self-dual parts must be of the same Petrov - Penrose type. So if this complex spacetime is a hyperheavenly space (with or without $\Lambda$) then by the Goldberg - Sachs theorem it admits both self-dual and anti-self-dual congruences of null strings which intersect each other and these intersections constitute the congruence of null geodesics. To assume the existence of null Killing vector field and identify this field with congruence of null geodesics seems to be the natural first step in investigating Lorentzian slices.

Our paper is organized as follows.

In section 2 we investigate the general properties of Killing vectors, especially of null ones. Some useful theorems are given and connection between null Killing vector and null strings is pointed out. Then the detailed discussion on the possible Petrov - Penrose types admitting null Killing symmetry is presented. Section 3 is a concise summary of the properties of hyperheavenly spaces. The main goal of our work is to present explicit form of the metrics with null Killing symmetry. The results are gathered in sections 4 and 5. There are seven different hyperheavenly metrics with null isometric or homothetic Killing vector and five different heavenly metrics. In section 6 we discuss the possible real slices of the metrics found in preceding sections. Some two-sided Walker and globally Osserman spaces are obtained. The Lorentzian slices of the type [II] and [D] are found. Concluding remarks ends our paper.

\section{Null Killing vectors and null strings.}
\setcounter{equation}{0}

\subsection{Killing equations and their integrability conditions in spinorial formalism.}

The system of Killing equations are given by
\begin{equation}
\nabla_{(a} K_{b)} = \chi \, g_{ab}
\end{equation}
The Killing vector is said to be \textsl{conformal}, if $\chi \ne \textrm{const}$, \textsl{homothetic} if $\chi=\chi_0= \textrm{const} \ne 0$ and \textsl{isometric} if $\chi=0$. For our purposes it is useful to present the Killing equations and their integrability conditions in the spinorial formalism. Let $(e^{1}, e^{2}, e^{3}, e^{4})$ be a null tetrad and $(\partial_{1}, \partial_{2}, \partial_{3}, \partial_{4})$ its inverse basis. Then the respective spinorial images are given by
\begin{equation}
(g^{A\dot{B}}) := \sqrt{2}
\left[\begin{array}{cc}
e^4 & e^2 \\
e^1 & -e^3
\end{array}\right] 
\ \ \ \ \ \ \ 
(\partial_{A\dot{B}}) := -\sqrt{2}
\left[\begin{array}{cc}
\partial_{4} & \partial_{2} \\
\partial_{1} & -\partial_{3}
\end{array}\right] 
\end{equation}
We use the following rules of manipulating the spinorial indices
\begin{equation}
\label{spinorial_indices_lowering_rule}
m_{A} = \ \in_{A B} m^{B}
\ , \ \ \ 
m^{A} = m_{B} \in^{BA}
\ , \ \ \
m_{\dot{A}} = \ \in_{\dot{A} \dot{B}} m^{\dot{B}}
\ , \ \ \ 
m^{\dot{A}} = m_{\dot{B}} \in^{\dot{B} \dot{A}}
\end{equation}
where $\in_{AB}$ and $\in_{\dot{A}\dot{B}}$ are the spinor Levi-Civita symbols
\begin{eqnarray}
&& \in_{AB}  := \left[ \begin{array}{cc}
                            0 & 1   \\
                           -1 & 0  
                            \end{array} \right] =:  \in^{AB} 
\ \ \ , \ \ \ 
 \in_{\dot{A}\dot{B}}  := \left[ \begin{array}{cc}
                            0 & 1   \\
                           -1 & 0  
                            \end{array} \right] =:  \in^{\dot{A}\dot{B}}  
\\ \nonumber
&& \in_{AC} \in^{AB} = \delta^{B}_{C} \ \ \ , \ \ \ \in_{\dot{A}\dot{C}} \in^{\dot{A}\dot{B}} = \delta^{\dot{B}}_{\dot{C}} \ \ \ , \ \ \ 
\delta^{A}_{C}= \delta^{\dot{B}}_{\dot{C}}= \left[ \begin{array}{cc}
                            1 & 0   \\
                            0 & 1  
                            \end{array} \right]
\end{eqnarray}
Correspondence between the null tetrad formalism and spinorial formalism is realized with the use of the spin-tensor $g^{aA \dot{B}}$ which is defined by the relation $g^{A \dot{B}} = g_{a}^{\ A \dot{B}} \, e^{a}$. It is easy to see that $-\frac{1}{2} g^{aA\dot{B}}g_{bA\dot{B}} = \delta^{a}_{b}$ and $-\frac{1}{2}g^{aA\dot{B}}g_{aC\dot{D}} = \delta^{A}_{C} \delta^{\dot{B}}_{\dot{D}}$. The operators $\partial^{A\dot{B}}$ and $\nabla^{A\dot{B}}$ are the spinorial images of operators $\partial^{a}$ and $\nabla^{a}$, respectively, given by
\begin{equation}
\partial^{A\dot{B}} = g_{a}^{\  A \dot{B}} \partial^{a}  \ \ \ \ \ \ \ \ \ \ \ \ \ \ 
\nabla^{A\dot{B}} = g_{a}^{\  A \dot{B}} \nabla^{a} 
\end{equation}
A conformal Killing vector $K$ can be written as
\begin{equation}
\label{ogolna_postac_wektora_Killll}
K = K^{a} \, \partial_{a} = -\frac{1}{2} K_{A\dot{B}} \partial^{A\dot{B}} 
\end{equation}
Components $K^{a}$ and $K_{A \dot{B}}$ are related by 
\begin{equation}
K^{a} = - \frac{1}{2} g^{a A \dot{B}} \, K_{A \dot{B}} \ \ \Longleftrightarrow \ \ 
K_{A \dot{B}} = g_{a  A \dot{B}} \, K^{a}
\end{equation}
Conformal Killing equations with conformal factor $\chi$ in spinorial form read
\begin{equation}
\label{wyjsciowa_postac_ro_Killinga}
\nabla_{A}^{\ \; \dot{B}} K_{C}^{\ \dot{D}} + \nabla_{C}^{\ \; \dot{D}} K_{A}^{\ \dot{B}} = -4 \chi \in_{AC} \in^{\dot{B}\dot{D}}
\end{equation}
which is equivalent to the following system of equations
\begin{subequations}
\begin{eqnarray}
\label{rozklad_rownania_Killinga}
E_{AC}^{\ \ \ \dot{B}\dot{D}} &\equiv& \nabla_{(A}^{\ \; (\dot{B}} K_{C)}^{\ \; \dot{D})} = 0 
\\
\label{definicja_chi}
E &\equiv& \nabla^{N\dot{N}} K_{N\dot{N}} +8 \chi = 0
\end{eqnarray}
\end{subequations}
From (\ref{rozklad_rownania_Killinga}) and (\ref{definicja_chi}) it follows that
\begin{equation}
\label{ogolne_rownanie_Killinga_spinnorowo}
\nabla_{A}^{\ \; \dot{B}} K_{C}^{\ \; \dot{D}} = l_{AC} \in^{\dot{B}\dot{D}} + l^{\dot{B}\dot{D}} \in_{AC} - 2 \chi  \in_{AC} \in^{\dot{B}\dot{D}}
\end{equation}
with
\begin{equation}
\label{definicja_eli}
l_{AC} := \frac{1}{2} \nabla_{(A}^{\ \ \dot{N}} K_{C)\dot{N}} \ \ \ \ \ \ \ \ \ \ \ \ 
l^{\dot{B}\dot{D}} := \frac{1}{2} \nabla^{N(\dot{B}} K_{N}^{\ \; \dot{D})}
\end{equation}
The integrability conditions of (\ref{rozklad_rownania_Killinga}) and (\ref{definicja_chi}) in Einstein space ($C_{AB \dot{C}\dot{D}}=0$, $R=-4 \Lambda$) consist of the following equations
\begin{subequations}
\begin{eqnarray}
\label{integrability_L_undot}
L_{RST}^{\ \ \ \ \; \dot{A}} &\equiv& \nabla_{R}^{\ \; \dot{A}} l_{ST} + 2C^{N}_{\ RST} K_{N}^{\ \; \dot{A}} + \frac{2}{3} \Lambda \in_{R(S}K_{T)}^{\ \ \dot{A}} + 2 \in_{R(S} \nabla_{T)}^{\ \ \dot{A}} \chi =0 \ \ \ \ \ \ \ \ 
\\
\label{integrability_L_dot}
L_{\dot{R}\dot{S}\dot{T}}^{\ \ \ \ \; A} &\equiv& \nabla^{A}_{\ \dot{R}} l_{\dot{S}\dot{T}} 
+ 2C^{\dot{N}}_{\ \dot{R}\dot{S}\dot{T}} K_{\ \ \dot{N}}^{A}+ \frac{2}{3} \Lambda \, \in_{\dot{R} ( \dot{S}}K^{A}_{\ \ \dot{T})} + 2 \in_{\dot{R} ( \dot{S}} \nabla^{A}_{\ \ \dot{T})} \chi=0
\\
\label{integrability_M_undot}
M_{ABCD} &\equiv& K_{N\dot{N}} \nabla^{N\dot{N}}C_{ABCD} + 4C^{N}_{\ \; (ABC} l_{D)N} - 4 \chi C_{ABCD} = 0
\\
\label{integrability_M_dot}
M_{\dot{A}\dot{B}\dot{C}\dot{D}} &\equiv& K_{N\dot{N}} \nabla^{N\dot{N}}C_{\dot{A}\dot{B}\dot{C}\dot{D}} + 4C^{\dot{N}}_{\ \; (\dot{A}\dot{B}\dot{C}} l_{\dot{D})\dot{N}} - 4 \chi C_{\dot{A}\dot{B}\dot{C}\dot{D}} = 0
\\
\label{integrability_N}
N_{AB}^{\ \ \ \dot{A}\dot{B}} &\equiv& \nabla_{A}^{\ \; \dot{A}} \nabla_{B}^{\ \; \dot{B}} \chi - \frac{2}{3} \Lambda \chi \in_{AB} \in^{\dot{A}\dot{B}} =0
\\
\label{integrability_R_undot}
R_{ABC}^{\ \ \ \ \ \dot{A}} &\equiv& C^{N}_{\ \; ABC} \nabla_{N}^{\ \; \dot{A}} \chi =0
\\
\label{integrability_R_dot}
R_{\dot{A}\dot{B}\dot{C}}^{\ \ \ \ \ A} &\equiv& C^{\dot{N}}_{\ \; \dot{A}\dot{B}\dot{C}} \nabla^{A}_{\ \; \dot{N}} \chi = 0
\end{eqnarray}
\end{subequations}

\subsection{Null strings via null Killing vectors.}

The existence of a null Killing vector has a significant influence on the geometry of the space. To explain this we first note that the null Killing vector can be presented in the form
\begin{equation}
\label{rozklad_zerowego_Killinga}
K_{A \dot{B}} = \mu_{A} \nu_{\dot{B}} \ \ \ \Longleftrightarrow \ \ \ K_{A \dot{B}} K^{A \dot{B}} = 0
\end{equation}
where $\mu_{A}$ and $\nu_{\dot{B}}$ are some nonzero spinors.
\newline
Moreover, it is well known that every spinor symmetric in all indices can be decomposed according to the formula
\begin{equation} 
\Psi_{A_{1}A_{2}...A_{n}} = \Psi_{(A_{1}A_{2}...A_{n})} = \Psi^{(1)}_{(A_{1}} \Psi^{(2)}_{A_{2}}... \Psi^{(n)}_{A_{n})}
\end{equation}
where $\Psi^{(i)}_{A}$ are some basic spinors. In particular, there exist spinors $\mathcal{A}_{A}$, $\mathcal{B}_{A}$, $\mathcal{A}_{\dot{A}}$ and $\mathcal{B}_{\dot{A}}$ such that
\begin{equation}
\label{rozklad_spinorow_lAB}
l_{AB} = \mathcal{A}_{(A} \mathcal{B}_{B)} \ \ \ , \ \ \ l_{\dot{A}\dot{B}} = \mathcal{A}_{(\dot{A}} \mathcal{B}_{\dot{B})}
\end{equation}
We prove the following
\newline
\newline
\textbf{Lemma 2.1}
\newline
Spinors $l_{AB}$ and $l_{\dot{A}\dot{B}}$ can be brought to the form $l_{AB}=\mu_{(A} \mathcal{B}_{B)}$ and $l_{\dot{A}\dot{B}}=\nu_{(\dot{A}} \mathcal{B}_{\dot{B})}$ without any loss of generality.
\newline
\newline
\textbf{Proof}
\newline
Inserting (\ref{rozklad_zerowego_Killinga}) and (\ref{rozklad_spinorow_lAB}) into (\ref{ogolne_rownanie_Killinga_spinnorowo}) we obtain
\begin{equation}
\label{rownanie_Killinga_po_wstawieniu}
\mu_{C} \nabla_{A}^{\ \ \dot{B}} \nu^{\dot{D}} + \nu^{\dot{D}} \nabla_{A}^{\ \ \dot{B}} \mu_{C} = 
\mathcal{A}_{(A} \mathcal{B}_{C)} \in^{\dot{B}\dot{D}} + \mathcal{A}^{(\dot{B}} \mathcal{B}^{\dot{D})} \in_{AC} - 2\chi \in_{AC}  \in^{\dot{B}\dot{D}}
\end{equation}
Contracting (\ref{rownanie_Killinga_po_wstawieniu}) with $\mu^{A}\mu^{C}\nu_{\dot{D}}$ one gets
\begin{equation}
\nu^{\dot{B}} \, \mathcal{A}_{A}\mu^{A} \, \mathcal{B}_{C}\mu^{C}=0
\end{equation}
so $\mathcal{A}_{A}$ or $\mathcal{B}_{A}$ must be proportional to $\mu_{A}$. Let $\mathcal{A}_{A}=\mathcal{A} \mu_{A}$, $\mathcal{A} \ne 0$. Re-defining spinor $\mathcal{B}_{A}$ (absorbing $\mathcal{A}$ into $\mathcal{B}_{A}$) we finally get $l_{AB} = \mu_{(A} \mathcal{B}_{B)}$. Analogously we prove that $l_{\dot{A}\dot{B}} = \nu_{(\dot{A}} \mathcal{B}_{\dot{B})}$. $\blacksquare$
\newline
\newline
\textbf{Theorem 2.2}
\newline
Let the null Killing vector $K_{A \dot{B}}$ be of the form (\ref{rozklad_zerowego_Killinga}). Then the two-dimensional self-dual holomorphic distribution $\{\mu_{A} \nu_{\dot{B}}, \mu_{A} \rho_{\dot{B}} \}$, $\nu_{\dot{B}} \rho^{\dot{B}} \ne 0$, is integrable and its integral manifolds constitute the congruence of self-dual null strings and the anti-self-dual distribution $\{\mu_{A}\nu_{\dot{B}}, \sigma_{A}\nu_{\dot{B}} \}$, $\mu_{A} \sigma^{A} \ne 0$, is also integrable and its integral manifolds constitute the congruence of anti-self-dual null strings. Moreover, both Weyl spinors $C_{ABCD}$ and $C_{\dot{A}\dot{B}\dot{C}\dot{D}}$ are algebraically special with $\mu_{A}$ and $\nu_{\dot{B}}$ being the undotted and dotted, respectively, multiple Penrose spinors.
\newline
\newline
\textbf{Proof}
\newline
Contracting (\ref{rownanie_Killinga_po_wstawieniu}) with $\mu^{A}\mu^{C}$ and remembering that $\mathcal{A}_{A}=\mu_{A}$ we get
\begin{equation}
\mu^{B} \mu^{C} \, \nabla_{B}^{\ \ \dot{A}} \mu_{C} = 0
\end{equation}
This means that the spinor $\mu_{A}$ defines a congruence of self-dual null strings in the sense that the 2-dimensional holomorphic distribution $\{\mu_{A} \nu_{\dot{B}}, \mu_{A} \rho_{\dot{B}} \}$, $\nu_{\dot{A}} \rho^{\dot{A}} \ne 0$ is integrable and its integrable manifolds constitute the congruence of self-dual null strings. From the complex Sachs-Goldberg theorem it follows, that $C_{ABCD}$ is algebraically special and $\mu_{A}$ is multiple Penrose dotted spinor, i.e.
\begin{equation}
C_{ABCD} \mu^{A}\mu^{B}\mu^{C}=0
\end{equation}
Analogously we prove that
\begin{equation}
\nu^{\dot{B}} \nu^{\dot{C}} \, \nabla^{A}_{\ \; \dot{B}} \nu_{\dot{C}} = 0 
\ \ \ \ \stackrel{\textrm{Goldberg Sach theorem}}{\Longleftrightarrow} \ \ \ \ C_{\dot{A}\dot{B}\dot{C}\dot{D}} \nu^{\dot{A}}\nu^{\dot{C}} \nu^{\dot{C}} = 0
\end{equation}
$\blacksquare$
\newline
\newline
In particular from Theorem 2.2 it follows that the integral curves of a null Killing vector are given by the intersection of self-dual and anti-self-dual congruences of null strings. 
\newline
Note that
\begin{eqnarray}
\label{struny}
\mu^{B} \mu^{C} \, \nabla_{B}^{\ \ \dot{A}} \mu_{C} = 0 & \Longleftrightarrow & 
\nabla_{B}^{\ \ \dot{A}} \mu_{C} = Z_{B}^{\ \ \dot{A}} \mu_{C}  + \in_{BC}  \Theta^{\dot{A}}
\\ \nonumber
\nu^{\dot{B}} \nu^{\dot{C}} \, \nabla^{A}_{\ \; \dot{B}} \nu_{\dot{C}} = 0 &\Longleftrightarrow&
\nabla^{A}_{\ \; \dot{B}} \nu_{\dot{C}} = X^{A}_{\ \ \dot{B}} \nu_{\dot{C}} + \in_{\dot{B}\dot{C}}  \Theta^{A}
\end{eqnarray}
where $\Theta^{A}$ and $\Theta^{\dot{A}}$ describe the optic properties of the anti-self-dual and self-dual null strings, respectively. Indeed, if $\Theta^{\dot{A}}=0$ then the self-dual null strings are  parallely-propagated, if $\Theta^{A}=0$ then anti-self-dual null strings are parallely propagated. Inserting (\ref{struny}), $\mathcal{A}_{A}=\mu_{A}$, $\mathcal{A}_{\dot{A}}=\nu_{\dot{A}}$ into (\ref{rownanie_Killinga_po_wstawieniu}), after some straightforward calculations we obtain
\begin{equation}
X_{A\dot{B}}=-Z_{A\dot{B}} \ \ \ , \ \ \ \mathcal{B}_{A}=\Theta_{A} \ \ \ , \ \ \ \mathcal{B}_{\dot{A}} = \Theta_{\dot{A}} \ \ \ , \ \ \ \mu_{A}\Theta^{A} + \nu_{\dot{A}} \Theta^{\dot{A}} + 4 \chi=0
\end{equation}
Let us prove another important theorem.
\newline
\newline
\textbf{Theorem 2.3}
\newline
Assume, that at least one of the $C_{ABCD}$ or $C_{\dot{A}\dot{B}\dot{C}\dot{D}}$ is nonzero. Then
\begin{eqnarray}
\nonumber
(i) \ \ \ \ \textrm{if } \Lambda \ne 0 \  \ & \textrm{then} & \  \ \chi=0
\\
\nonumber
(ii) \ \ \ \ \textrm{if } \Lambda = 0 \  \ & \textrm{then} & \  \ \chi=\chi_{0} = \textrm{const} 
\end{eqnarray}
\textbf{Proof}
\newline
Assume, that $C_{ABCD} \ne 0$. Then from (\ref{integrability_R_undot}) it follows, that that $\nabla_{A}^{\ \; \dot{A}} \chi$ is the quadruple Debever-Penrose spinor. However, as is well known, two quadruple DP spinors are necessarily lineary dependent so $\nabla^{A \dot{1}} \chi$ has to be proportional to $\nabla^{A \dot{2}} \chi$ or, equivalently
\begin{equation}
\label{one_quadruple_DP_spinor_condition}
\nabla_{A\dot{A}} \chi  \cdot   \nabla^{A\dot{A}} \chi =0
\end{equation}
Acting on (\ref{one_quadruple_DP_spinor_condition}) with $\nabla_{B}^{\ \; \dot{B}}$ and using (\ref{integrability_N}) one quickly obtains
\begin{equation}
\Lambda \chi \, \nabla_{B}^{\ \; \dot{B}} \chi =0
\end{equation}
Hence if $\Lambda \ne 0$ then $\nabla_{B}^{\ \; \dot{B}} \chi =0$. Finally, using (\ref{integrability_N})) we get $\chi =0$ what proves $(i)$.
\newline
If $\Lambda=0$, then still $\nabla_{N}^{\ \ \dot{A}} \chi$ is a quadruple DP-spinor. However, we proved that $\mu_{N}$ is a multiple DP-spinor, so it must be
\begin{equation}
\label{rozklad_pochodnej_nablachi}
\nabla_{N\dot{A}} \chi = \mu_{N} \chi_{\dot{A}}
\end{equation}
with some $\chi_{\dot{A}}$. Inserting (\ref{rozklad_pochodnej_nablachi}) into (\ref{integrability_N}) and contracting with $\mu^{B}$ we arrive at the conclusion $\Theta^{\dot{A}}\chi^{\dot{B}}=0$, so if we want to maintain possible conformal symmetries, the self-dual null string defined by the (conformal) Killing vector must be nonexpanding, $\Theta^{\dot{A}}=0$. Consequently $l_{\dot{A}\dot{B}}=0$. Inserting this into (\ref{integrability_L_dot}) and contracting it with $\in^{\dot{R}\dot{S}}$ we finally get $\chi^{\dot{A}}=0$. From (\ref{rozklad_pochodnej_nablachi}) it follows that $\nabla_{N\dot{A}} \chi=0$ and this proves $(ii)$. $\blacksquare$
\newline
\newline
Summing up, null conformal symmetries can appear only in the Einstein spaces with $C_{ABCD}=0=C_{\dot{A}\dot{B}\dot{C}\dot{D}}$, i.e. in the de-Sitter space (with $\Lambda \ne 0$) or in Minkowski space (with $\Lambda =0$). We do not consider these spaces here.

The null Killing vector field defines congruence of null (complex) geodesics. The optical properties of Killing vector field can be easily obtained. One gets
\begin{subequations}
\begin{eqnarray}
\label{geodesic_expansion}
\textrm{expansion} &:=& \frac{1}{2} \nabla^{a}K_{a} = 2\chi_{0}
\\
\label{geodesic_twist}
\textrm{twist}^{2} &:=& \frac{1}{2} \nabla_{[a}K_{b]} \, \nabla^{[a}K^{b]} = -2 \chi_{0}^{2}
\\
\label{geodesic_shear}
\textrm{shear} \cdot \dot{\textrm{shear}} &:=& \frac{1}{2} \nabla_{(a}K_{b)} \, \nabla^{(a}K^{b)} - \frac{1}{4} \big( \nabla^{a}K_{a} \big)^{2} = -2 \chi_{0}^{2}
\end{eqnarray}
\end{subequations}
Thus we conclude, that null homothetic Killing field defines null geodesic congruence with nonzero expansion, twist and shear, while null isometric Killing field is nonexpanding, nontwisting and shearfree.

Gathering above considerations: we reduced the problem of null Killing vectors in Einstein space to the set of equations
\begin{subequations}
\begin{eqnarray}
\textrm{form of the Killing vector}: && K_{A \dot{B}} = \mu_{A}\nu_{\dot{B}} 
\\
\textrm{spinors } l_{AB} \textrm{ and } l_{\dot{A}\dot{B}}: &&
l_{AB} = \mu_{(A}\Theta_{B)} \ \ \ , \ \ \ l_{\dot{A}\dot{B}} = \nu_{(\dot{A}} \Theta_{\dot{B})}
\\
\textrm{homothetic factor:} &&
\chi_{0}=\textrm{const} \ \ \ ,  \ \ \ \Lambda \chi_{0} = 0
\\ 
\label{rownania_strun_po_oblicz}
\textrm{self-dual null string:} &&\nabla_{A \dot{B}} \mu_{C} = Z_{A \dot{B}} \mu_{C}  + \in_{AC}  \Theta_{\dot{B}}
\\ 
\label{rownania_strun_po_oblicz_antiselfdual}
\textrm{anti-self-dual null string:} &&
\nabla_{A \dot{B}} \nu_{\dot{C}} = -Z_{A \dot{B}} \nu_{\dot{C}} + \in_{\dot{B}\dot{C}}  \Theta_{A} \ \ \ \ \ 
\\ 
\label{rownanie_Killinga_ktorezostalo}
\textrm{Killing equation:} &&\mu_{A}\Theta^{A} + \nu_{\dot{A}} \Theta^{\dot{A}} + 4 \chi_{0}=0
\end{eqnarray}
\end{subequations}
Algebraic degeneration conditions $C_{ABCD}\mu^{A}\mu^{B}\mu^{C}=0$ and $C_{\dot{A}\dot{B}\dot{C}\dot{D}}\nu^{\dot{A}}\nu^{\dot{B}}\nu^{\dot{C}}=0$ can be combined with (\ref{integrability_L_undot}) and (\ref{integrability_L_dot}). After some work we obtain
\begin{subequations}
\begin{eqnarray}
\label{algebraiczna_degeneracja_undot}
&&2C^{N}_{\ RST} \mu_{N} + \Sigma \, \mu_{R}\mu_{S}\mu_{T} + (3 \Omega +\Lambda) \mu_{(R}\mu_{s}m_{T)}  = 0
\ \  \ \ \textrm{with} \ \ m_{A}\mu^{A}=1 \ \ \ 
\\
\label{algebraiczna_degeneracja_dot}
&&2C^{\dot{N}}_{ \ \dot{R}\dot{S}\dot{T}} \nu_{\dot{N}} + \dot{\Sigma} \, \nu_{\dot{R}} \nu_{\dot{S}} \nu_{\dot{T}} + (3\dot{\Omega} + \Lambda) \nu_{(\dot{R}} \nu_{\dot{S}} n_{\dot{T})}=0
\ \ \ \ \ \ \, \textrm{with} \ \ n_{\dot{A}} \nu^{\dot{A}} = 1 \ \ \ 
\end{eqnarray}
\end{subequations}
where $\Sigma$, $\dot{\Sigma}$, $\Omega$ and $\dot{\Omega}$ are defined by the relations
\begin{subequations}
\begin{eqnarray}
\label{ostateczny_L_undot}
&&\nabla_{R}^{\ \ \dot{A}} \big( \mu_{(S}\Theta_{T)} \big) = \nu^{\dot{A}} \big( \Sigma \, \mu_{R}\mu_{S}\mu_{T} + 2 \Omega \, \mu_{R} \mu_{(S}m_{T)} + (\Lambda + \Omega) \mu_{S}\mu_{T}m_{R} \big)
\\ 
\label{ostateczny_L_dot}
&&\nabla_{R}^{\ \ \dot{A}} \big( \nu_{(\dot{S}} \Theta_{\dot{T})}  \big) = \mu_{R} \big( \dot{\Sigma} \, \nu^{\dot{A}} \nu_{\dot{S}}\nu_{\dot{T}} + 2 \dot{\Omega} \, \nu^{\dot{A}} \nu_{(\dot{S}}n_{\dot{T})} + (\Lambda + \dot{\Omega}) n^{\dot{A}} \nu_{\dot{S}} \nu_{\dot{T}} \big)
\end{eqnarray}
\end{subequations}
We end this subsection by pointing two relations, essential in further analysis. Contracting (\ref{ostateczny_L_undot}) with $\mu_{S}\mu_{T}$ and using (\ref{rownania_strun_po_oblicz}) we obtain 
\begin{equation}
\label{pierwsze_wazne}
\Theta^{\dot{A}} \mu_{S}\Theta^{S}=0
\end{equation}
Analogously, contracting (\ref{ostateczny_L_dot}) with $\nu_{\dot{S}}\nu_{\dot{T}}$ then using  (\ref{rownania_strun_po_oblicz_antiselfdual}) we obtain  
\begin{equation}
\label{drugie_wazne}
\Theta^{{A}} \nu_{\dot{S}}\Theta^{\dot{S}}=0
\end{equation}
Now we are ready to discuss the possible algebraic types admitting null Killing vector.

\subsection{Null homothetic symmetries.}

Here we assume $\chi_{0} \ne 0$, what immediately gives $\Lambda =0$. Simple analysis of equations (\ref{pierwsze_wazne}) - (\ref{drugie_wazne}) together with (\ref{rownanie_Killinga_ktorezostalo}) bring us to the conlcusion, that the only possibilities are
\begin{itemize}
\item $\Theta^{\dot{A}}=0$ (self-dual null string is nonexpanding), $\mu_{A}\Theta^{A} \ne 0$ (anti-self-dual null string is necesarilly expanding, more even, expansion $\Theta^{A}$ cannot be proportional to DP-spinor $\mu^{A}$)
\item $\Theta^{A}=0$ (anti-self-dual null string is nonexpanding), $\nu_{\dot{A}}\Theta^{\dot{A}} \ne 0$ (self-dual null string is necesarilly expanding, more even, expansion $\Theta^{\dot{A}}$ cannot be proportional to DP-spinor $\nu^{\dot{A}}$)
\end{itemize}
Of course, both possibilities constitute Eintein spaces with the same geometric properties. It is enough to consider only one of them with detailes, say $\Theta^{\dot{A}}=0$. From (\ref{ostateczny_L_dot}) we conclude, that $\dot{\Sigma}=\dot{\Omega}=0$. Careful analysis of (\ref{ostateczny_L_undot}) gives $\Omega=0$. From (\ref{algebraiczna_degeneracja_undot}) and (\ref{algebraiczna_degeneracja_dot}) we obtain
\begin{subequations}
\begin{eqnarray}
\label{zmienione_L_ne_symetriehom}
&&2C^{N}_{\ RST} \mu_{N} + \Sigma \, \mu_{R}\mu_{S}\mu_{T} =0
\\
&&2C^{\dot{N}}_{ \ \dot{R}\dot{S}\dot{T}} \nu_{\dot{N}} =0 
\end{eqnarray}
\end{subequations}
where
\begin{equation}
\label{definicja_Sigmy_dla_homotetii}
\Sigma \, \mu_{R}\mu_{T} \nu_{\dot{A}} := \nabla_{R \dot{A}} \Theta_{T} + \Theta_{T} Z_{R \dot{A}} 
\end{equation}
(the last formula is a consequence of (\ref{ostateczny_L_undot}).

The only possible anti-self-dual Petrov types are $[\textrm{N},-]$. From (\ref{zmienione_L_ne_symetriehom}) we easily get, that the only possible self-dual Petrov types are $[\textrm{III},-]$. Self-dual type $[\textrm{N}]$ is not admitted. Indeed, assume, that $C_{ABCD}$ is of the type $[\textrm{N}]$, so $C^{N}_{\ RST} \mu_{N}=0$, what gives $\Sigma=0$. Contracting (\ref{definicja_Sigmy_dla_homotetii}) with $\Theta^{T}$ we obtain $\Theta^{T} \, \nabla_{R \dot{A}} \Theta_{T}=0$, so $\Theta^{T}$ defines the congruence of the self-dual null strings. But we proved earlier (see Theorem 2.2), that self-dual null string is defined by $\mu^{T}$. The number of independent congruences of self-dual null strings is equal the number of multiple undotted DP-spinors, so there are infinitely many independent congruences of self-dual null strings in the heavenly spaces, two in the self-dual type $[\textrm{D}]$ and only one in the self-dual types $[\textrm{II}, \textrm{III}, \textrm{N}]$. But here we examine self-dual type $[\textrm{N}]$, so there is only one congruence of the null-strings. It means, that $\Theta^{T}$ must be proportional to $\mu^{T}$ or $\Theta^{T}\mu_{T}=0 \ \rightarrow \chi_{0}=0$. This contradicts our assumption, that $\chi_{0} \ne 0$. It proves, that the only possible Petrov types which admitt null homothetic symmetries are $[\textrm{III},-] \otimes [\textrm{N},-]$. Self-dual null string is nonexpanding, anti-self-dual null string must be expanding. 

[Remark: considering the second possibility with $\Theta^{A}=0$ we obtain possible Petrov types $[\textrm{N},-] \otimes [\textrm{III},-]$, but still type $[\textrm{III}]$ corresponds to nonexpanding null string, and the type $[\textrm{N}]$ corresponds to expanding null string].

All possible types via geometric properties of the null strings are presented in the table below:
\begin{displaymath}
\renewcommand{\arraystretch}{1.4}
\begin{tabular}{|c|c|c|}   \hline
 & self-dual null string is  
 & self-dual null string is                          
 \\ 
 & nonexpanding $\Theta^{\dot{A}} = 0$  
 & expanding $\Theta^{\dot{A}} \ne 0$                           
 \\ \hline
anti-self-dual null string is & not admitted & $[\textrm{N},-]^{e} \otimes [\textrm{III},-]^{n}$
\\ 
nonexpanding $\Theta^{A} = 0$ &  & 
\\ \hline
anti-self-dual null string is & $[\textrm{III},-]^{n} \otimes [\textrm{N},-]^{e}$ & not admitted
\\ 
expanding $\Theta^{A} \ne 0$ & &
\\ \hline
\end{tabular}
\end{displaymath}

All independent possibilities are given in detailes in the subsections:
\renewcommand{\arraystretch}{1.4}
\begin{displaymath}
\begin{tabular}{|c|c|c|c|}   \hline
 Type & SD null string & ASD null string  & subsection
\\ \hline
\multicolumn{4}{|c|}{hyperheavenly metrics }
\\ \hline
$[\textrm{N}]^{e} \otimes [\textrm{III}]^{n}$ & expanding & nonexpanding  & \ref{subsection_nullhomothetic_hyperheavenly}
\\ \hline
\multicolumn{4}{|c|}{heavenly metrics }
\\ \hline
$[\textrm{N}]^{e} \otimes [-]^{n}$ & expanding & nonexpanding  & \ref{subsection_nullhomothetic_heavenly_N}
\\
$[\textrm{III}]^{n} \otimes [-]^{e}$ & nonexpanding & expanding  &  \ref{subsection_nullhomothetic_heavenly_III}
\\ \hline
\end{tabular}
\end{displaymath}
[The upper index $e$ means, that the corresponding null string is expanding, index $n$ - nonexpanding].

\subsection{Null isometric symmetries.}

Here we assume $\chi_{0}=0$. Analysis of equations (\ref{pierwsze_wazne}) - (\ref{drugie_wazne}) and (\ref{rownanie_Killinga_ktorezostalo}) proves, that $\Theta^{A}$ and $\Theta^{\dot{A}}$ must have the form
\begin{equation}
\label{form_ofthe_Theta_in_isometry}
\Theta^{A} = \Theta \, \mu^{A} \ \ \ \ , \ \ \ \ \Theta^{\dot{A}} = \dot{\Theta} \, \nu^{\dot{A}}
\end{equation}
so
\begin{equation}
\label{form_ofthe_lAB_in_isometry}
l_{AB} = \Theta \, \mu_{A} \mu_{B} \ \ \ \ , \ \ \ \ l_{\dot{A}\dot{B}} = \dot{\Theta} \, \nu_{\dot{A}} \nu_{\dot{B}}
\end{equation}
Using (\ref{form_ofthe_Theta_in_isometry}) and (\ref{form_ofthe_lAB_in_isometry}) in the (\ref{ostateczny_L_undot}) and (\ref{ostateczny_L_dot}) we find $\Omega=\dot{\Omega}=\Theta \dot{\Theta}$. Equations (\ref{algebraiczna_degeneracja_undot}) and (\ref{algebraiczna_degeneracja_dot}) can be rearranged to the form
\begin{subequations}
\begin{eqnarray}
\label{curvatura_integra_undot}
&&2C^{N}_{\ RST} \mu_{N} + \Sigma \, \mu_{S}\mu_{R}\mu_{T} + (\Lambda +3 \Theta \dot{\Theta}) \, \mu_{(S}\mu_{R}m_{T)}=0
\\
\label{curvatura_integra_dot}
&&2C^{\dot{N}}_{\ \dot{R}\dot{S}\dot{T}} \nu_{\dot{N}} + \dot{\Sigma} \, \nu_{\dot{S}}\nu_{\dot{R}}\nu_{\dot{T}} + (\Lambda +3 \Theta \dot{\Theta}) \, \nu_{(\dot{S}}\nu_{\dot{R}}n_{\dot{T})}=0
\end{eqnarray}
\end{subequations}
and the equations (\ref{ostateczny_L_undot}) and (\ref{ostateczny_L_dot}) read
\begin{subequations}
\begin{eqnarray}
\label{definition_Sigma}
&&\nabla_{R \dot{A}} \Theta + 2\Theta \, Z_{R\dot{A}} =: \nu_{\dot{A}} \big( \Sigma \mu_{R} + (\Lambda + 3\Theta \dot{\Theta}) m_{R} \big)  
\\
\label{definition_Sigmadot}
&&\nabla_{R\dot{A}} \dot{\Theta} - 2 \dot{\Theta} \, Z_{R\dot{A}} =: \mu_{R} \big( \dot{\Sigma} \nu_{\dot{A}} + (\Lambda + 3\Theta \dot{\Theta}) n_{\dot{A}} \big)  
\end{eqnarray}
\end{subequations}
Multiplying (\ref{definition_Sigma}) by $\dot{\Theta}$ and (\ref{definition_Sigmadot}) by $\Theta$ and adding both equations one arrives to the useful formula
\begin{equation}
\label{warunek_przy_wyrugowanym_Z}
\frac{1}{3} \, \nabla_{R \dot{A}} (\Lambda + 3\Theta \dot{\Theta}) = (\Sigma \dot{\Theta} + \dot{\Sigma}\Theta) \mu_{R} \nu_{\dot{A}} + (\Lambda + 3 \Theta \dot{\Theta}) (\Theta \, \mu_{R} n_{\dot{A}} + \dot{\Theta} \, m_{R} \nu_{\dot{A}})
\end{equation}
When both null strings are nonexpanding ($\Theta=\dot{\Theta}=0$) then from (\ref{definition_Sigma}) and (\ref{definition_Sigmadot}) it follows that $\Lambda=\Sigma=\dot{\Sigma}=0$. Consequently, from (\ref{curvatura_integra_undot}) and (\ref{curvatura_integra_dot}) we obtain that the only possible types are 
$[\textrm{N},-] \otimes [\textrm{N},-]$.

If anti-self-dual null string is nonexpanding ($\Theta=0$) and the self-dual null string is expanding ($\dot{\Theta} \ne 0$) then from (\ref{definition_Sigma}) we get $\Lambda=\Sigma=0$, so the self-dual type is at most of the type $[\textrm{N}]$. The anti-self-dual type can be of the type $[\textrm{III}]$ (if $\dot{\Sigma} \ne 0$) or of the type $[\textrm{N},-]$ (if $\dot{\Sigma} =0$), so in this case we deal with the types $[\textrm{N},-] \otimes [\textrm{III,N},-]$. [The case with expanding anti-self-dual string and nonexpanding self-dual string has the same geometry and leads to the types $[\textrm{III,N},-] \otimes [\textrm{N},-]$. 

In the last case both null strings are expanding $\Theta \ne 0$, $\dot{\Theta} \ne 0$. Equations (\ref{curvatura_integra_undot}) and (\ref{curvatura_integra_dot}) give in general types $[\textrm{II,D}] \otimes [\textrm{II,D}]$. Cosmological constant $\Lambda$ can be arbitrary here. [It does not follow from (\ref{curvatura_integra_undot}) - (\ref{curvatura_integra_dot}), but the mixt types $[\textrm{II}] \otimes [\textrm{D}]$ and $[\textrm{D}] \otimes [\textrm{II}]$ are not admitted, we will prove it during further analysis]. 

Now deal with the self-dual type $[\textrm{III}]$, $C_{ABCD}\mu^{C}\mu^{D}=0$. From (\ref{curvatura_integra_undot}) we got $\Lambda + 3\Theta \dot{\Theta}=0$ (so the cosmological constant $\Lambda$ is necessarily nonzero), from (\ref{curvatura_integra_dot}) we conclude, that the anti-self-dual type is $[\textrm{III}]$. Indeed, anti-self-dual types $[\textrm{N},-]$ extort $\dot{\Sigma}=0$, what combined with (\ref{warunek_przy_wyrugowanym_Z}) gives $\Sigma=0$. But $\Sigma=0$ automatically reduces the self-dual type to $[\textrm{N},-]$. 

In self-dual types $[\textrm{N},-]$, $C_{ABCD}\mu^{D}=0$ so $\Sigma=0=\Lambda + 3\Theta \dot{\Theta}$. Immediatelly we have $\dot{\Sigma}=0$, so anti-self-dual type is $[\textrm{N},-]$ too. Like in previous case, cosmological constant $\Lambda$ must be nonzero here. 

All possible types are gathered in the table below:
\begin{displaymath}
\renewcommand{\arraystretch}{1.4}
\begin{tabular}{|c|c|c|}   \hline
 & self-dual null string is  
 & self-dual null string is                          
 \\ 
 & nonexpanding $ \dot{\Theta} = 0$  
 & expanding $\dot{\Theta} \ne 0$                           
 \\ \hline
anti-self-dual null string is & $[\textrm{N},-] \otimes [\textrm{N},-], \Lambda=0$ & $[\textrm{N},-] \otimes [\textrm{III,N},-], \Lambda=0$
\\ 
nonexpanding $\Theta = 0$ &  & 
\\ \hline
anti-self-dual null string is & $[\textrm{III,N},-] \otimes [\textrm{N},-], \Lambda=0$ & $[\textrm{II}] \otimes [\textrm{II}],[\textrm{D}] \otimes [\textrm{D}], \Lambda$ arbitrary
\\ 
expanding $\Theta \ne 0$ & & $[\textrm{III}] \otimes [\textrm{III}], \Lambda \ne 0$
\\
 & & $[\textrm{N},-] \otimes [\textrm{N},-], \Lambda \ne 0$
\\ \hline
\end{tabular}
\end{displaymath}
\newline
\newline
All independent metrics are presented in detailes in subsections
\begin{displaymath}
\renewcommand{\arraystretch}{1.4}
\begin{tabular}{|c|c|c|c|c|}   \hline
 Type & SD null string & ASD null string & $\Lambda$ & subsection
\\ \hline
\multicolumn{5}{|c|}{hyperheavenly metrics }
\\ \hline
$[\textrm{N}]^{n} \otimes [\textrm{N}]^{n}$ & nonexpanding & nonexpanding & 0 &
\ref{subsection_nullisometric_hyperheavenly_NxN_nn}
\\
$[\textrm{III,N}]^{n} \otimes [\textrm{N}]^{e}$ & nonexpanding & expanding & 0 &
\ref{subsection_nullisometric_hyperheavenly_III_NxN}
\\
$[\textrm{II}]^{e} \otimes [\textrm{II}]^{e}$ & expanding & expanding & arbitrary & \ref{subsection_nullisometric_hyperheavenly_IIxII}
\\
$[\textrm{D}]^{e} \otimes [\textrm{D}]^{e}$ & expanding & expanding & arbitrary & \ref{subsection_nullisometric_hyperheavenly_DxD}
\\
$[\textrm{III}]^{e} \otimes [\textrm{III}]^{e}$ & expanding & expanding & $\ne 0$ &
\ref{subsection_nullisometric_hyperheavenly_IIIxIII}
\\
$[\textrm{N}]^{e} \otimes [\textrm{N}]^{e}$ & expanding & expanding & $\ne 0$ &
\ref{subsection_nullisometric_hyperheavenly_IIIxIII}
\\ \hline
\multicolumn{5}{|c|}{heavenly metrics }
\\ \hline
$[\textrm{N}]^{n} \otimes [-]^{n}$ & nonexpanding & nonexpanding & 0 &
\ref{subsection_nullisometric_hyperheavenly_NxN_nn}
\\
$[\textrm{III,N}]^{n} \otimes [-]^{e}$ & nonexpanding & expanding & 0 &
\ref{subsection_nullisometric_hyperheavenly_III_NxN}
\\
$[\textrm{N}]^{e} \otimes [-]^{n}$ & expanding & nonexpanding & 0 &
\ref{subsection_nullisometric_hyperheavenly_III_NxN}
\\
$[\textrm{N}]^{e} \otimes [-]^{e}$ & expanding & expanding & $\ne 0$ &
\ref{subsection_nullisometric_hyperheavenly_IIIxIII}
\\ \hline
\end{tabular}
\end{displaymath}

\section{Hyperheavenly spaces.}
\setcounter{equation}{0}

The considerations from the previous section allow to establish all possible algebraic types of the spaces, which admit the null Killing symmetry. Main aim of present paper is to present the explicit metrics with such symmetries. Due to Theorem 2.2, null Killing vector defines the congruence of both self-dual and anti-self-dual null strings and extorts the algebraic degeneration of both self-dual and anti-self-dual part of the Weyl curvature spinor. Let us remind the definition of \textsl{hyperheavenly space}.
\newline
\newline
\textbf{Definition 3.1}
\newline
\textsl{ Hyperheavenly space ($\mathcal{HH}$-space) with cosmological constant} is a 4 - dimensional complex analytic differential manifold endowed with a holomorphic Riemannian metric $ds^{2}$ satisfying the vacuum Einstein equations with cosmological constant and such that the self - dual or anti - self - dual part of the Weyl tensor is algebraically degenerate. These kind of spaces admits a congruence of totally null, self-dual (or anti-self-dual, respectively) surfaces. $\blacksquare$

In general, hyperheavenly spaces require only self-dual (or anti-self-dual) congruences of null strings. 
The spaces which admitt the null Killing vector are equipped in both self-dual and anti-self-dual congruences of null strings, so they are automaticaly hyperheavenly spaces. In hyperheavenly spaces, vacuum Einstein equations can be reduced to one, nonlinear, partial differential equation of the second order, for one holomorphic function. This equation is called \textsl{hyperheavenly equation}. It seems, that using the hyperheavenly formalism in order to obtain the explicite metrics which admitt the null Killing vector is most natural. 

The existence of the null strings allows to introduce some useful tetrad and the coordinate system. The self-dual null string generated by the null Killing vector is given by the equation $\mu^{B} \mu^{C} \, \nabla_{B}^{\ \ \dot{A}} \mu_{C} = 0$ which is equivalent to the Pfaff system $\mu_{A} g^{A \dot{B}}=0$. Choosing the spinorial basis in such a manner, that $\mu_{A}=(0, \mu_{2})$, $\mu_{2} \ne 0$ we arrive to conclusion, that the self-dual null string is defined by the Pfaff system
\begin{equation}
e^{1}=0 \ \ \ , \ \ \ e^{3}=0 \ \ \ \Leftrightarrow \ \ \ g^{2 \dot{A}}=0
\end{equation}
(the surface element of the null string is given by $e^{1} \wedge e^{3}$).
\newline
A null tetrad $(e^{1},e^{2},e^{3},e^{4})$ and a coordinate system $(q_{\dot{A}}, p^{\dot{B}})$ can be always chosen so that
\begin{eqnarray}
\label{tetrada_spinorowa}
-\frac{1}{\sqrt{2}} \, g^{2}_{\ \, \dot{A}} = \left[ \begin{array}{c}  e^{3}  \\ e^{1}  \end{array} \right] &=& \phi^{-2} \, dq_{\dot{A}}
\\ \nonumber
\frac{1}{\sqrt{2}} \, g^{1 \dot{A}} = \left[ \begin{array}{c}  e^{4}  \\ e^{2}  \end{array} \right]  &=&
-dp^{\dot{A}} + Q^{\dot{A} \dot{B}} \, dq_{\dot{B}}
\end{eqnarray}
where $\phi$ and $Q^{\dot{A} \dot{B}}=Q^{\dot{B} \dot{A}}$ are holomorphic functions. Coordinates $q_{\dot{A}}$ label the null strings, hence $p^{\dot{A}}$ are coordinates on them. Dual basis is given by
\begin{equation}
\label{baza_dualna}
-\partial_{\dot{A}} = 
\left[ \begin{array}{c}  \partial_{4}  \\ \partial_{2}  \end{array} \right] 
 \ , \ \ \  
\eth^{\dot{A}} = 
\left[ \begin{array}{c}  \partial_{3}  \\ \partial_{1}  \end{array} \right] 
\  , \ \ \  
\partial^{A \dot{B}} = \sqrt{2} (\delta_{1}^{A} \eth^{\dot{B}} - \delta_{2}^{A} \partial^{\dot{B}} )
\end{equation}
where
\begin{equation}
\partial_{\dot{A}} := \frac{\partial}{\partial p^{\dot{A}}} \ , \ \ \ 
\eth^{\dot{A}} := \phi^{2} \left( \frac{\partial}{\partial q_{\dot{A}}} + Q^{\dot{A} \dot{B}} \partial_{\dot{B}} \right)
\end{equation}
Of course, for consistency with (\ref{spinorial_indices_lowering_rule}), the rules to raise and lower spinor indices in spinorial differential operators read $\partial^{\dot{A}} = \partial_{\dot{B}} \in^{\dot{A} \dot{B}}$, $\partial_{\dot{A}} = \in_{\dot{B} \dot{A}} \partial^{\dot{B}}$, $\eth^{\dot{A}} = \eth_{\dot{B}} \in^{\dot{A} \dot{B}}$, $\eth_{\dot{A}} = \in_{\dot{B} \dot{A}} \eth^{\dot{B}}$, so
\begin{equation}
\partial^{\dot{A}} = \frac{\partial}{\partial p_{\dot{A}}}
\ , \ \ \ 
\eth_{\dot{A}} = \phi^{2} \left( \frac{\partial}{\partial q^{\dot{A}}} - Q_{\dot{A}}^{\ \ \dot{B}} \partial_{\dot{B}} \right)
\end{equation}
The metric $ds^2$ is given by
\begin{equation}
\label{metryka_ogolna}
ds^{2} =2 \, e^{1} \underset{s}{\otimes} e^{2} + 2 \, e^{3} \underset{s}{\otimes} e^{4}
= -\frac{1}{2} \, g_{A\dot{B}} \underset{s}{\otimes} g^{A\dot{B}}
= 2 \phi^{-2} \, (- dp^{\dot{A}} \underset{s}{\otimes} dq_{\dot{A}} + 
           Q^{\dot{A} \dot{B}} \, dq_{\dot{A}} \underset{s}{\otimes} dq_{\dot{B}})
\end{equation}

The congruence of null strings have some invariant properties. Investigating the equation $\nabla_{B}^{\ \ \dot{A}} \mu_{C} = Z_{B}^{\ \ \dot{A}} \mu_{C}  + \in_{BC}  \Theta^{\dot{A}}$ with $\mu_{1}=0$, $\mu_{2} \ne 0$ we easily find, that
\begin{eqnarray}
\Theta_{\dot{A}} &=& \mathbf{\Gamma}_{112\dot{A}} \, \mu_{2}
\\
Z_{A\dot{B}} &=& -\mathbf{\Gamma}_{12A \dot{B}} - \in_{A2} \mathbf{\Gamma}_{112\dot{B}} + \partial_{A \dot{B}} \ln \mu_{2}
\end{eqnarray}
If $\Theta^{\dot{A}}=0 \ \Longleftrightarrow \ \mathbf{\Gamma}_{112\dot{A}}=0$ then self-dual null strings are parallely propagated. The hyperheavenly spaces based on such null strings are called \textsl{nonexpanding}. If the null string is not parallely propagated ($\Theta^{\dot{A}} \ne 0 \ \Longleftrightarrow \ \mathbf{\Gamma}_{112\dot{A}} \ne 0$), the corresponding hyperheavenly space is called \textsl{expanding}. 

Vacuum Einstein equations impose some constraints on $\phi$ and $Q^{\dot{A}\dot{B}}$. The final forms of the $\phi$ and $Q^{\dot{A}\dot{B}}$ are esentially different in expanding and nonexpanding hyperheavenly spaces.
\newline
\newline
\textbf{Nonexpanding hyperheavenly spaces}
\newline
The reduction of Einstein equations brings us to
\begin{equation}
\label{funkcje_metryki_dla_nonexxpand}
\phi=1  \ \ \ \ , \ \ \ \ 
Q^{\dot{A} \dot{B}} = -  \Theta_{p_{\dot{A}} p_{\dot{B}}} 
+ \frac{2}{3} F^{(\dot{A}} p^{\dot{B})} +\frac{1}{3} \Lambda \, p^{\dot{A}} p^{\dot{B}}
\end{equation}
Einstein equations can be reduced to \textsl{nonexpanding hyperheavenly equation with $\Lambda$}
\begin{eqnarray}
\frac{1}{2} \, \Theta_{p_{\dot{A}} p_{\dot{B}}} \Theta_{p^{\dot{A}} p^{\dot{B}}}
+ \Theta_{p_{\dot{A}} q^{\dot{A}}} 
+F^{\dot{A}} \Big( \Theta_{p^{\dot{A}}} - \frac{2}{3} \, p^{\dot{B}} \, \Theta_{p^{\dot{A}} p^{\dot{B}}}  \Big) + \frac{1}{18} \, (F^{\dot{A}} p_{\dot{A}})^{2}&&
\\ \nonumber
+ \frac{1}{6} \, \frac{\partial F_{\dot{A}}}{\partial q^{\dot{B}}} \, p^{\dot{A}} p^{\dot{B}} 
+\Lambda \Big( p^{\dot{A}} \Theta_{p^{\dot{A}}} - \Theta - \frac{1}{3} \, p^{\dot{A}} p^{\dot{B}} \Theta_{p^{\dot{A}} p^{\dot{B}}}   \Big)
&=& N_{\dot{A}} \, p^{\dot{A}} + \gamma \ \ \ \ \ \ \ \ \ \ \ \ \
\end{eqnarray}
where $F^{\dot{A}}$, $N^{\dot{A}}$ and $\gamma$ are arbitrary functions of $q^{\dot{C}}$ only (constant on each null string), $\Lambda$ is a cosmological constant and $\Theta = \Theta (p^{\dot{A}}, q_{\dot{B}})$ is \textsl{the key function}. The metric is given by (\ref{metryka_ogolna}) with $\phi$ and $Q^{\dot{A}\dot{B}}$ in the form (\ref{funkcje_metryki_dla_nonexxpand}).
\newline
\newline
\textbf{Expanding hyperheavenly spaces}
\newline
If the congruence of self-dual null strings is expanding, we obtain
\begin{equation}
\phi=J_{\dot{A}}p^{\dot{A}}  \ \ \ \ , \ \ \ \ Q^{\dot{A} \dot{B}} = -2 \, J^{(\dot{A}} \partial^{\dot{B})} W - \phi \, \partial^{\dot{A}} \partial^{\dot{B}} W+ \frac{1}{\tau^{2}} K^{\dot{A}} K^{\dot{B}} \Big( \mu  \phi^{3} + \frac{\Lambda}{6}  \Big)
\end{equation}
where $\mu=\mu(q^{\dot{N}})$ is an arbitrary function, $\Lambda$ is the cosmological constant and $W=W(p^{\dot{A}}, q_{\dot{B}})$ is \textsl{the key function}. $J_{\dot{A}}$ and $K_{\dot{A}}$ are constant, nonzero spinors, connected by the relation
\begin{equation}
K^{\dot{A}} J_{\dot{B}} - K_{\dot{B}} J^{\dot{A}} = \tau \, \delta^{\dot{A}}_{\dot{B}}
\ \ \ \ \textrm{where} \ \ \ \ \tau = K^{\dot{A}} J_{\dot{A}} \ne 0
\end{equation}
[$\tau$ is an arbitrary constant, not loosing generality one can set $\tau=1$.]
\newline
Einstein equations can be reduced to \textsl{expanding hyperheavenly equation with $\Lambda$}
\begin{eqnarray}
\label{expanding_hyperheavenly_equation_form2}
&&\frac{1}{2} \phi^{4} (\phi^{-2}W_{p_{\dot{B}}})_{p_{\dot{A}}}
                     (\phi^{-2}W_{p^{\dot{B}}})_{p^{\dot{A}}}
+\phi^{-1} W_{p_{\dot{A}} q^{\dot{A}}} 
- \mu \phi^{4} \, [ \phi^{-1} (\phi^{-1}W)_{\phi} ]_{\phi}
\\ \nonumber
&&+ \frac{\eta}{2 \tau^{2}} \Big( \eta J^{\dot{C}} - \phi K^{\dot{C}} \Big) \mu_{q^{\dot{C}}} - \frac{\Lambda}{6} \, \phi^{-1} \, W_{\phi \phi}
= N_{\dot{A}} \, p^{\dot{A}} + \gamma
\end{eqnarray}
where $N^{\dot{A}}$ and $\gamma$ are arbitrary functions of $q^{\dot{C}}$ only (constant on each self-dual null string). Instead of the $(p^{\dot{A}}, q_{\dot{B}})$-coordinate system , another one, namely $(\phi, \eta, w,t )$ is universally used
\begin{eqnarray}
&& \phi=J_{\dot{A}} p^{\dot{A}} \ \ \ , \ \ \ \eta := K^{\dot{A}} p_{\dot{A}} \ \ \ \Longleftrightarrow \ \ \
 \tau  p^{\dot{A}} = \eta   J^{\dot{A}} + \phi  K^{\dot{A}} 
\\ \nonumber
&& w := J_{\dot{A}}q^{\dot{A}} \ \ \ , \ \ \ t := K^{\dot{A}}q_{\dot{A}}   \ \ \ \Longleftrightarrow \ \ \ 
\tau q_{\dot{A}} = t J_{\dot{A}} + w K_{\dot{A}}
\end{eqnarray}
with operators
\begin{equation}
\partial_{\phi} = \frac{1}{\tau} K^{\dot{A}} \partial_{\dot{A}} 
\ \ \ , \ \ \ 
\partial_{\eta} = \frac{1}{\tau} J^{\dot{A}} \partial_{\dot{A}} 
\ \ \ , \ \ \ 
\partial_{w} = \frac{1}{\tau} K^{\dot{A}} \frac{\partial}{\partial q^{\dot{A}}}
\ \ \ , \ \ \ 
\partial_{t} = \frac{1}{\tau} J^{\dot{A}} \frac{\partial}{\partial q^{\dot{A}}}
\end{equation}
In $(\phi, \eta, w,t)$-language, the hyperheavenly equation reads
\begin{eqnarray}
&& \tau^{2} \Big( W_{\eta \eta}W_{\phi \phi} - W_{\eta \phi}W_{\eta \phi} + 2 \phi^{-1} W_{\eta} W_{\eta \phi} - 2 \phi^{-1} W_{\phi}W_{\eta\eta} \Big) + \tau \phi^{-1} \Big( W_{w\eta}-W_{t\phi} \Big)
\ \ \ \ \ \ 
\\ \nonumber
&&- \mu \Big( \phi^{2} W_{\phi\phi} - 3 \phi W_{\phi} + 3 W \Big) + \frac{\eta}{2 \tau} (\mu_{t} \eta - \mu_{w} \phi) - \frac{\Lambda}{6} \phi^{-1} W_{\phi \phi} = \frac{1}{2} \varkappa \phi - \frac{1}{2} \nu \eta + \gamma
\end{eqnarray}
where $2N_{\dot{A}} =: \nu K_{\dot{A}} + \varkappa J_{\dot{A}}$.
\newline
The metric
\begin{eqnarray}
\label{metrykaprzestrzeni_HH_z_ekspanja}
ds^{2} &=& (\phi \tau)^{-2} \bigg\{ 2\tau (d \eta \underset{s}{\otimes} d w - d \phi \underset{s}{\otimes} dt) + 2 \, \Big( -\tau^{2} \phi \, W_{\eta\eta} + \mu \phi^{3}+\frac{\Lambda}{6}  \Big) 
dt \underset{s}{\otimes}dt \ \ \ \ \ \ 
\\ \nonumber
&& \ \ \ \ \ \ \ \ \ \ +4 \left( -\tau^{2} \phi \, W_{\eta \phi} + \tau^{2} \, W_{\eta}\right) dw \underset{s}{\otimes}dt
  + 2 \left(  -\tau^{2} \phi \, W_{\phi \phi} + 2\tau^{2} \, W_{\phi}         \right) dw \underset{s}{\otimes}dw \bigg\} \ \ \ \ \ \ 
\end{eqnarray}
We do not present here the curvature formulas and connection forms. Interested readers can found them in [...].
\newline
Using the connection forms, calculated explicitly in \cite{biblio_16} one can note, that in expanding hyperheavenly spaces expansion of the congruence of the self-dual null strings is proportional to nonzero spinor $J_{\dot{A}}$, namely $\Theta_{\dot{A}}  = -\sqrt{2} \phi^{-1} \mu_{2} \, J_{\dot{A}}$. Hence $\Theta_{\dot{A}} \ne 0$.

\section{Metrics admitting null homothetic symmetries.}
\setcounter{equation}{0}

\subsection{Hyperheavenly spaces of the type $[\textrm{N}]^{e} \otimes [\textrm{III}]^{n}$}
\label{subsection_nullhomothetic_hyperheavenly}

The hyperheavenly spaces which admit the null homothetic symmetry must be of types $[\textrm{N}]^{e} \otimes [\textrm{III}]^{n}$ or $[\textrm{III}]^{n} \otimes [\textrm{N}]^{e}$. Such spaces have been considered previously in \cite{biblio_16} and \cite{biblio_dod_3} but without any detailes. Since the type $[\textrm{III}]^{n} \otimes [\textrm{N}]^{e}$ can be obtained from $[\textrm{N}]^{e} \otimes [\textrm{III}]^{n}$ just by changing the orientation we study the case $[\textrm{N}]^{e} \otimes [\textrm{III}]^{n}$ and use the general formulas from \cite{biblio_16}. Killing vector has the form
\begin{equation}
\label{Killing_vector_for_III_N_homot}
K=-2 \chi_{0} p^{\dot{A}} \frac{\partial}{\partial p^{\dot{A}}}
\end{equation}
and
\begin{equation}
\label{Properties_Killing_vector_for_III_N_homot}
l_{AB}=0 \  , \ \ \ l^{\dot{A}\dot{B}} = 4\chi_{0} \phi^{-1} \, J^{(\dot{A}}p^{\dot{B})}
\end{equation}
The key function and the curvature reads
\begin{eqnarray}
\label{funkcja_kluczowa_dla_hyperheavenly_homotheticnull}
 && W=\phi^{2} F(x, w, t) \ , \ \ \  x:=\frac{\eta}{\phi}
\\ 
\label{krzywizna_dla_hyperheavenly_homotheticnull}
 && C^{(1)} = 2\phi^{7} \,\tau \gamma_{t} \  , \ \ \  C_{\dot{A}\dot{B}\dot{C}\dot{D}} = C_{(\dot{A}}p_{\dot{B}}p_{\dot{C}} p_{\dot{D})}
\\ \nonumber
&& C_{\dot{A}} := 4 \tau^{3} \phi^{-2} F_{xxx} J_{\dot{A}} + \tau^{4} \phi^{-3} F_{xxxx} p_{\dot{A}}
\end{eqnarray}
where $F=F(x,w,t)$ and $\gamma=\gamma(w,t)$ are arbitrary functions of their arguments, such that $F_{xxx} \ne 0$ and $\gamma_{t} \ne 0$. Inserting the key function $W$ (\ref{funkcja_kluczowa_dla_hyperheavenly_homotheticnull}) into hyperheavenly equation we get
\begin{equation}
\label{hyperheavenlyeq_nullhomothetic}
F_{x}^{2} - 2F F_{xx} + F_{xw'} + xF_{xt'} - 2F_{t'} = \gamma'
\end{equation}
where $w'=\tau w$, $t'=\tau t$ oraz $\gamma'=\tau^{-2} \gamma$.  The general solution of the equation (\ref{hyperheavenlyeq_nullhomothetic}) is not known. The metric has the form (remember, that $\eta=x\phi$) 
\begin{eqnarray}
\label{metryka_homote_typpu_NxIII}
ds^{2} &=& 2\phi^{-2} \bigg\{ \tau^{-1} (d \eta \underset{s}{\otimes} d w - d \phi \underset{s}{\otimes} dt) -   \phi \, F_{xx}  \ dt \underset{s}{\otimes}dt \ \ \ \ \ \ 
\\ \nonumber
&& \ \ \ \ \ \ \ \ \ \ +2  \phi \,  x F_{xx} \ dw \underset{s}{\otimes}dt
  +   \phi  \left( 2F-x^{2} F_{xx}    \right) dw \underset{s}{\otimes}dw \bigg\} \ \ \ \ \ \ 
\end{eqnarray}
where $F=F(x,w,t)$ satisfies the equation (\ref{hyperheavenlyeq_nullhomothetic}).

\subsection{Heavenly spaces of the type $[\textrm{N}]^{e} \otimes [-]^{n}$}
\label{subsection_nullhomothetic_heavenly_N}

There are two, essentially different heavenly reductions of the hyperheavenly space of the type $[\textrm{N}]^{e} \otimes [\textrm{III}]^{n}$ with null homothetic symmetry. Taking $F_{xxx}=0 \Longleftrightarrow C_{\dot{A}} = 0 \Longleftrightarrow C_{\dot{A}\dot{B}\dot{C}\dot{D}}=0$ in (\ref{krzywizna_dla_hyperheavenly_homotheticnull}) we obtain the space of the type $[\textrm{N}]^{e} \otimes [-]^{n}$ with expanding self-dual null string. 

Equation (\ref{hyperheavenlyeq_nullhomothetic}) under the additional assumption $F_{xxx}=0$ can be easily solved. Using gauge freedom, which is still available (see \cite{biblio_16} for detailes) one gets
\begin{eqnarray}
 && W=\frac{1}{2} f(t,w) \, \phi^{2}
\\ 
 && C^{(1)} = -2\phi^{7} \, f_{tt} 
\end{eqnarray}
where $f=f(w,t)$ is an arbitrary function. The metric is
\begin{equation}
\label{metryka_homote_typpu_Nxnic}
ds^{2}= \frac{2}{\tau} \phi^{-2} (d\eta \underset{s}{\otimes} dw - d\phi \underset{s}{\otimes} dt 
             + \tau f \phi \, dw \underset{s}{\otimes} dw )
\end{equation}

\subsection{Heavenly spaces of the type $[\textrm{III}]^{n} \otimes [-]^{e}$}
\label{subsection_nullhomothetic_heavenly_III}

The second possible heavenly reduction of the hyperheavenly space of the type $[\textrm{N}]^{e} \otimes [\textrm{III}]^{n}$ with null homothetic symmetry is the heavenly space of the type $[-]^{e} \otimes [\textrm{III}]^{n}$. Formally it is enough to set $C^{(1)}=0 \Longleftrightarrow \gamma_{t}=0$ in subsection \ref{subsection_nullhomothetic_hyperheavenly}. However, equation (\ref{hyperheavenlyeq_nullhomothetic}) still is hard to solve. It appeared to be much more convenient to attack the problem from the opposite side and consider the space of the type $[\textrm{III}]^{n} \otimes [-]^{e}$. As a starting point we take the nonexpanding hyperheavenly spaces and we set 
$C_{\dot{A}\dot{B}\dot{C}\dot{D}}=0$. It allows us to take the general key function as a third-order polynomial in $p^{\dot{A}}$ coordinates (see \cite{biblio_15,biblio_dod_3} for detailes). The metric appeared to be two-sided Walker \cite{biblio_9}. Using general results from \cite{biblio_dod_3} and \cite{biblio_9} (especially Theorem 5.1 from \cite{biblio_9}) we find the form of the Killing vector
\begin{equation}
K=2 \chi_{0} p^{\dot{A}} \frac{\partial}{\partial p^{\dot{A}}}
\end{equation}
and the spinors $l_{AB}$ and $l_{\dot{A}\dot{B}}$
\begin{equation}
l_{AB} = -4 \chi_{0} \delta^{1}_{(A} \delta^{2}_{B)}   \  , \ \ \ l_{\dot{A}\dot{B}}=0
\end{equation}
The key function and the curvature
\begin{eqnarray}
 && \Theta = \frac{1}{6} p^{\dot{1}}p^{\dot{M}} \bigg( p^{\dot{2}} \frac{\partial X}{\partial q^{\dot{M}}} - p^{\dot{1}}e^{X} \frac{\partial Y}{\partial q^{\dot{M}}} \bigg)
\\
 && 
C^{(2)}=2 \frac{\partial}{\partial q_{\dot{1}}} \bigg( e^{X} \frac{\partial Y}{\partial q_{\dot{1}}} - \frac{\partial X}{\partial q_{\dot{2}}} \bigg)
\ , \ \  C^{(1)} = p^{\dot{A}} \bigg( F_{\dot{A}} C^{(2)} + \frac{\partial C^{(2)}}{\partial q^{\dot{A}}} \bigg) \ \ \ \ \ \
\\ \nonumber
&&\textrm{where } F_{\dot{1}} = \frac{\partial X}{\partial q_{\dot{2}}} - 2 e^{X} \frac{\partial Y}{\partial q_{\dot{1}}}
\ \  , \ \  \ \ 
F_{\dot{2}} = \frac{\partial X}{\partial q_{\dot{1}}}
\end{eqnarray}
where $X=X(q^{\dot{M}})$ and $Y=Y(q^{\dot{M}})$ are arbitrary functions. The metric
\begin{eqnarray}
\label{metryka_homote_typpu_IIIxnic}
&&ds^{2}=2 \, (- dp^{\dot{A}} \underset{s}{\otimes} dq_{\dot{A}} + 
           Q^{\dot{A} \dot{B}} \, dq_{\dot{A}} \underset{s}{\otimes} dq_{\dot{B}}) 
\\ 
\nonumber
&& 
Q^{\dot{1}\dot{1}} = - p^{\dot{1}} \frac{\partial X}{\partial q^{\dot{2}}} \ \ \ ,  \ \ \
Q^{\dot{1}\dot{2}} = -p^{\dot{1}} e^{X} \frac{\partial Y}{\partial q^{\dot{2}}} \ \ \ , \ \ \ 
Q^{\dot{2}\dot{2}} = -p^{\dot{2}} \frac{\partial X}{\partial q^{\dot{1}}} - p^{\dot{2}} e^{X} \frac{\partial Y}{\partial q^{\dot{2}}} + p^{\dot{1}} e^{X} \frac{\partial Y}{\partial q^{\dot{1}}}
\ \ \ \ \ 
\end{eqnarray}
[Note, that the conditions $C_{\dot{A}\dot{B}\dot{C}\dot{D}}=0=\Lambda$ assure the infinitely many congruences of nonexpanding anti-self-dual null strings, that is why the space considered belongs to the two-sided Walker class. However, the anti-self-dual null string generated by the null Killing vector is expanding.]

\section{Metrics admitting null isometric symmetries.}
\setcounter{equation}{0}

\subsection{Spaces of the type $[\textrm{II}]^{e} \otimes [\textrm{II}]^{e}$}
\label{subsection_nullisometric_hyperheavenly_IIxII}

\subsubsection{General case $\Lambda \ne 0$}

If both null strings are expanding then null isometric symmetries are admitted by the types $[\textrm{II}] \otimes [\textrm{II}]$ and $[\textrm{D}] \otimes [\textrm{D}]$. The respective metrics have been discussed in \cite{biblio_dod_2} (with $\Lambda=0$) and then in \cite{biblio_16} (with $\Lambda \ne 0$). The Killing vector reads
\begin{equation}
K=\frac{\partial}{\partial \eta}
\end{equation}
and
\begin{equation}
l_{AB}= \frac{2\phi}{\tau} \bigg( \mu_{0} \phi^{3} + \frac{\Lambda}{6} \bigg) \delta_{(A}^{2}\delta_{B)}^{2} 
\  , \ \ \ l^{\dot{A}\dot{B}} = - \frac{2}{\tau} \phi^{-1} J^{\dot{A}} J^{\dot{B}}
\end{equation}
The key function and the curvature are 
\begin{eqnarray}
\label{funkcja_kluczowa_dla_IIxII}
 &&  W=W(\phi, w, t)
\\ 
\label{krzywizna_dla_hyperheavenly_isometricnull_IIxII}
 &&   C^{(3)} = -2 \mu_{0} \phi^{3} \  , \ \ \   
C^{(1)} = 6\tau\phi^{7} \mu_{0} W_{t}  \ , \ \ \ \mu_{0} = \textrm{const} \ne 0 \ \ \ \ \ \ 
\\ \nonumber
&& C_{\dot{A}\dot{B}\dot{C}\dot{D}} = -\phi^{3} \, J_{(\dot{A}}J_{\dot{B}}C_{\dot{C}}D_{\dot{D})}
 \ , \ \ \ 
C_{(\dot{C}}D_{\dot{D})} = \frac{6\mu_{0}}{\tau^{2}} K_{\dot{C}}K_{\dot{D}} - W_{\phi \phi \phi \phi} J_{\dot{C}}J_{\dot{D}}
\end{eqnarray}
The metric reads
\begin{eqnarray}
\label{metrykaprzestrzeni_IIxDD_oggolna}
ds^{2} &=& 2 \phi^{-2} \bigg\{ \tau^{-1} (d \eta \underset{s}{\otimes} d w - d \phi \underset{s}{\otimes} dt) +  \, \Big( \frac{\mu_{0}}{\tau^{2}} \phi^{3}+\frac{\Lambda}{6\tau^{2}}  \Big) 
dt \underset{s}{\otimes}dt \ \ \ \ \ \ 
\\ \nonumber
&& \ \ \ \ \ \ \ \ \ \ +  \left(  2 \, W_{\phi}  - \phi \, W_{\phi \phi}  \right) dw \underset{s}{\otimes}dw \bigg\} \ \ \ \ \ \ 
\end{eqnarray}
After inserting the key function (\ref{funkcja_kluczowa_dla_IIxII}) into the hyperheavenly equation we get
\begin{equation}
\label{typ_IIxII_hyperheavenly_equation}
\bigg( \mu_{0} \phi^{3} + \frac{\Lambda}{6} \bigg) W_{\phi \phi} - 3\mu_{0} \phi^{2} W_{\phi} +3\mu_{0} \phi W + \tau W_{t \phi} = 0
\end{equation}
In order to maintain the type $[\textrm{II}] \otimes [\textrm{II}]$ we have to assume $W_{t} \ne 0$ and $W_{\phi\phi\phi\phi} \ne 0$. The general solution of the equation (\ref{typ_IIxII_hyperheavenly_equation}) is not known. Its reduction to canonical form is realized by the transformation
\begin{equation}
\label{definicja_nowych_wspol}
s:= \frac{\mu_{0}}{\tau} \, t \ \ \ , \ \ \ z:= \frac{\mu_{0}}{\tau} \, t - \int \frac{d \phi}{\phi^{3} +\frac{\Lambda}{6 \mu_{0}} }
\end{equation}
Considering the key function $W$ as a function of the variables $(z,s,w)$ we obtain the equation
\begin{equation}
\label{typ_IIxII_hyperheavenly_equation_red1}
W_{zs} - 6 \phi^{2} W_{z} -3\phi \bigg( \phi^{3} + \frac{\Lambda}{6 \mu_{0}} \bigg) W=0
\end{equation}
Multiplying the equation (\ref{typ_IIxII_hyperheavenly_equation_red1}) by $\phi \big( \phi^{3} + \frac{\Lambda}{6 \mu_{0}} \big)^{-2}$ one can bring it to the form
\begin{equation}
\label{typ_IIxII_hyperheavenly_equation_red2}
\partial_{s} \bigg[ \phi \bigg( \phi^{3} + \frac{\Lambda}{6 \mu_{0}} \bigg)^{-2} W_{z} \bigg] - \partial_{z} \bigg[ \bigg( \phi^{3} + \frac{\Lambda}{6 \mu_{0}} \bigg)^{-1} W \bigg] =0
\end{equation}
From (\ref{typ_IIxII_hyperheavenly_equation_red2}) we infer the existence of the potential $\Sigma = \Sigma (z,s,w)$ such that
\begin{subequations}
\begin{eqnarray}
\label{potencjal_1}
&&\Sigma_{z} = \phi \bigg( \phi^{3} + \frac{\Lambda}{6 \mu_{0}} \bigg)^{-2} W_{z}
\\
\label{potencjal_2}
&&\Sigma_{s} = \bigg( \phi^{3} + \frac{\Lambda}{6 \mu_{0}} \bigg)^{-1} W
\end{eqnarray}
\end{subequations}
From the (\ref{potencjal_2}) one can calculate $W$
\begin{equation}
W=\bigg( \phi^{3} + \frac{\Lambda}{6 \mu_{0}} \bigg) \Sigma_{s}
\end{equation}
Inserting it into (\ref{potencjal_1}) we arrive to the equation
\begin{equation}
\label{typ_IIxII_hyperheavenly_equation_red3}
\Sigma_{zs} = 3\phi^{2} \Sigma_{s} + \phi^{-1} \bigg( \phi^{3} + \frac{\Lambda}{6 \mu_{0}} \bigg) \Sigma_{z}
\end{equation}
Of course, $\phi$ has to be considered as a function of coordintates $s$ and $z$, $\phi = \phi (z,s)$, according to (\ref{definicja_nowych_wspol}). The integral in (\ref{definicja_nowych_wspol}) can be calculated in all subcases, leading to the condition $z=z(s, \phi)$. Unfortunatelly, the inverse function $\phi = \phi (z,s)$ is an elementary function only if $\Lambda=0$. However, if $\Lambda=0$ more efficient transformation can be proposed.

\subsubsection{Special case $\Lambda = 0$}

In what follows we assume $\Lambda=0$ and we introduce the transformation $(\phi, \eta, w,t) \rightarrow (x,y,u,v)$:
\begin{equation}
\label{transformacja_lorentzowska_typu_IIxII}
\phi^{2} = \Big( - \frac{1}{\mu_{0}} \Big)^{\frac{2}{3}} \frac{1}{4x} \ , \ \ \ \frac{t}{\tau} = \Big( - \frac{1}{\mu_{0}} \Big)^{\frac{1}{3}} (x + iy) \ , \ \ \ \frac{w}{\tau} = -\frac{1}{4} \Big( - \frac{1}{\mu_{0}} \Big)^{\frac{2}{3}} u \ , \ \ \ \eta=v
\end{equation}
In coordinates $(x,y,u,v)$ the metric (\ref{metrykaprzestrzeni_IIxDD_oggolna}) has the form
\begin{equation}
\label{metryka_typu_IIxII_wpostaci_lorr}
ds^{2} = -2x \, du \underset{s}{\otimes} (dv +M du) + x^{-\frac{1}{2}} (dx \underset{s}{\otimes} dx + dy \underset{s}{\otimes} dy)
\end{equation}
where 
\begin{equation}
\label{definicja_function_M}
M := - \frac{\tau^{2}}{4} \Big( - \frac{1}{\mu_{0}} \Big)^{\frac{2}{3}} (2 W_{\phi} - \phi W_{\phi \phi})
\equiv 2\tau^{2} \Big( - \frac{1}{\mu_{0}} \Big)^{\frac{1}{3}} (\partial_{x} + i \partial_{y}) \big( x^{\frac{5}{2}} (\partial_{x} + i \partial_{y}) W \big)
\end{equation}
Obviously, function $M$ is the function of variables $(x,y,u)$. The hyperheavenly equation (\ref{typ_IIxII_hyperheavenly_equation}) takes the form
\begin{equation}
\label{hiperniebianskie_dla_typu_IIxII_formalizm_XY}
4x^{2} (W_{xx} + W_{yy}) +12x (W_{x} + iW_{y}) +3W =0
\end{equation}
differentiating equation (\ref{hiperniebianskie_dla_typu_IIxII_formalizm_XY}) twice by $\partial_{x} + i \partial_{y}$ and using definition of $M$, after some algebraic work we obtain
\begin{equation}
\label{rownannie_typu_IIxII_wpostaci_lorr}
xM_{xx} + xM_{yy} + M_{x} = 0
\end{equation}
Equation (\ref{rownannie_typu_IIxII_wpostaci_lorr}) is equivalent to the Euler-Poisson-Darboux equation (EPD equation) and its solutions have been discussed in literature \cite{biblio_dod_11}. The form of the metric (\ref{metryka_typu_IIxII_wpostaci_lorr}) and the equation (\ref{rownannie_typu_IIxII_wpostaci_lorr}) are especially useful in obtaining the Lorentzian slices (see section 6).

\subsection{Spaces of the type $[\textrm{D}]^{e} \otimes [\textrm{D}]^{e}$}
\label{subsection_nullisometric_hyperheavenly_DxD}

From the previous subsection one can easily obtain the general metric for the type $[\textrm{D}]^{e} \otimes [\textrm{D}]^{e}$ with $\Lambda$. Self-dual type $[\textrm{D}]$ we get after setting $C^{(1)}=0 \ \Rightarrow \ W_{t}=0 \ \Rightarrow \ W=W(\phi,w)$. General solution of the hyperheavenly equation (\ref{typ_IIxII_hyperheavenly_equation}) reads $W=\mu_{0} f_{1}(w) \phi^{3} + f_{2}(w) \phi - (\Lambda / 3) f_{1}(w)$ and it automatically causes anti-self-dual type beeing of the type $[\textrm{D}]$. Moreover, it can be proved that arbitrary functions $f_{1}$ and $f_{2}$ can be gauged to zero without any loss of generality. [We do not prove that fact here, but it can be easily done by using the results from \cite{biblio_16} together with some straightforward calculations]. Finally it brings us to conclusion, that in the hyperheavenly spaces of the type $[\textrm{D}] \otimes [\textrm{D}]$ admitting null Killing vector the key function $W$  can be gauge to zero. [The inverse implication is also true; putting $W=0$ in expanding hyperheavenly equation and all curvature formulas we easily get, that such hyperheavenly space automatically becomes of the type $[\textrm{D}] \otimes [\textrm{D}]$ with the metric (\ref{metryka_typu_DxD})]. The metric
\begin{equation}
\label{metryka_typu_DxD}
ds^{2} = (\phi \tau)^{-2} \bigg\{ 2\tau (d \eta \underset{s}{\otimes} d w - d \phi \underset{s}{\otimes} dt) + 2 \, \bigg( \mu_{0} \phi^{3}+\frac{\Lambda}{6}  \bigg) 
dt \underset{s}{\otimes}dt \bigg\}
\end{equation}
admitts, together with null isometric Killing vector $\partial_{\eta}$, three other isometric Killing vectors $\partial_{w}$, $\partial_{t}$ and $w\partial_{w} - \eta \partial{_\eta}$ and - if $\Lambda=0$ - one homothetic Killing vector $\frac{2}{3} \chi_{0} (2t\partial_{t} - \phi \partial_{\phi} + \eta \partial_{\eta})$.

If $\Lambda=0$ the metric (\ref{metryka_typu_DxD}) can be easily transformed to coordinate system $(x,y,u,v)$ defined by (\ref{transformacja_lorentzowska_typu_IIxII}). According to (\ref{definicja_function_M}) function $M=0$ and the metric reads
\begin{equation}
ds^{2} = -2x \, du \underset{s}{\otimes} dv  + x^{-\frac{1}{2}} (dx \underset{s}{\otimes} dx + dy \underset{s}{\otimes} dy)
\end{equation}

\subsection{Spaces of the type $[\textrm{III}]^{e} \otimes [\textrm{III}]^{e}$ and $[\textrm{N},-]^{e} \otimes [\textrm{N},-]^{e}$}
\label{subsection_nullisometric_hyperheavenly_IIIxIII}

The next possible metrics generated by null Killig vector with both self-dual and anti-self-dual null strings expanding, are metrics of the type $[\textrm{III}]^{e} \otimes [\textrm{III}]^{e}$, $[\textrm{N}]^{e} \otimes [\textrm{N}]^{e}$ and $[\textrm{N}]^{e} \otimes [-]^{e}$ or $[-]^{e} \otimes [\textrm{N}]^{e}$. All of them have been found in \cite{biblio_16}. Here we have the Killing vector 
\begin{equation}
K=\frac{\partial}{\partial \eta}
\end{equation}
and
\begin{equation}
l_{AB}=  \frac{\Lambda \phi}{3\tau} \, \delta_{(A}^{2}\delta_{B)}^{2} 
\  , \ \ \ l^{\dot{A}\dot{B}} = - \frac{2}{\tau} \phi^{-1} J^{\dot{A}} J^{\dot{B}}
\end{equation}
The key function and the curvature
\begin{eqnarray}
 &&  W= \alpha_{0} \eta \phi^{3} + f ( z,w ) - \frac{3}{7 \Lambda} \tau^{2} \alpha_{0}^{2} \, \phi^{7} + g_{t} \, \phi^{2} - \frac{\Lambda}{3\tau} \, g \, \phi \ , \ \ \  z:= \phi- \frac{\Lambda t}{6\tau}
\\ \nonumber
 &&     
C^{(2)} = -2 \tau \alpha_{0} \Lambda \, \phi^{5} \ , \ \ \ C^{(1)}=-4 \tau^{2} \phi^{7} \big( g_{ttt} + \alpha_{0}^{2} \Lambda \, \phi^{3} \big)
\\ \nonumber
&& C_{\dot{A}\dot{B}\dot{C}\dot{D}} = \phi^{3} \, J_{(\dot{A}}J_{\dot{B}}J_{\dot{C}}L_{\dot{D})}
\\ \nonumber
&& L_{\dot{A}} := \Big(  f_{zzzz} - \frac{360}{\Lambda} \tau^{2} \alpha_{0}^{2} \, \phi^{3} \Big) J_{\dot{A}} - 24 \alpha_{0} \, K_{\dot{A}}
\end{eqnarray}
where $f=f(z,w)$ and $g=g(w,t)$ are arbitrary functions of their variables. The metric
\begin{eqnarray}
\label{metryka_typu_IIIxIII}
ds^{2} &=& (\phi \tau)^{-2} \bigg\{ 2\tau (d \eta \underset{s}{\otimes} d w - d \phi \underset{s}{\otimes} dt) +   \frac{\Lambda}{3}  \, dt \underset{s}{\otimes}dt  -8 \tau^{2} \alpha_{0} \phi^{3} \, dw \underset{s}{\otimes}dt
\\ \nonumber
&& \ \ \ \ \ \ \ \ \ \ 
  + 2 \Big(  -\tau^{2}  f_{zz} \, \phi+ 2 \tau^{2} f_{z} + \frac{12}{\Lambda} \tau^{4} \alpha_{0}^{2} \, \phi^{6} + 2 \tau^{2} g_{t} \, \phi - \frac{2}{3}\tau \Lambda g    \Big) dw \underset{s}{\otimes}dw \bigg\} \ \ \ \ \ \ 
\end{eqnarray}
In all formulas $\Lambda \ne 0$. Particular types are characterized by
\begin{itemize}
\item type $[\textrm{III}]^{e} \otimes [\textrm{III}]^{e}$: $\alpha_{0} \ne 0$, $\alpha_{0}$ can be re-gauged to $1$ without any loss of generality
\item type $[\textrm{N}]^{e} \otimes [\textrm{N}]^{e}$: $\alpha_{0}=0$, $g_{ttt} \ne 0$, $f_{zzzz} \ne 0$
\item type $[\textrm{N}]^{e} \otimes [-]^{e}$: $\alpha_{0}=0$, $g_{ttt} \ne 0$, $f_{zzzz}=0$ (one can set $f=0$ without any loss of generality)
\item type $[-]^{e} \otimes [\textrm{N}]^{e}$: $\alpha_{0}=0$,  $f_{zzzz} \ne 0$, $g_{ttt}=0$ (one can set $g=0$ without any loss of generality)
\end{itemize}
Of course, spaces of the types $[\textrm{N}]^{e} \otimes [-]^{e}$ and $[-]^{e} \otimes [\textrm{N}]^{e}$ have the same geometry of null strings, since both self-dual and anti-self-dual null strings are expanding. The case $[-]^{e} \otimes [\textrm{N}]^{e}$ has been considered with detailes in \cite{biblio_dod_10}. It describes the general heavenly metric with $\Lambda$ admitting null Killing vector. In real case this solution has the signature $(++--)$ and it is general metric of the 4-dimensional global Osserman space with non-zero curvature scalar admitting the null Killing vector.

\subsection{Spaces of the type $[\textrm{III,N,}-]^{n} \otimes [\textrm{N,}-]^{e}$}
\label{subsection_nullisometric_hyperheavenly_III_NxN}

In this case we deal with the hyperheavenly types $[\textrm{III,N}]^{n} \otimes [\textrm{N}]^{e}$ with nonexpanding self-dual null string defined by the null Killing vector. Anti-self-dual null string is still expanding. These metrics have been discussed in \cite{biblio_dod_3}. The Killing vector has the form
\begin{equation}
K=q^{\dot{1}} \frac{\partial}{\partial p^{\dot{1}}}
\end{equation}
Then
\begin{equation}
l_{AB}=  - \delta_{(A}^{2}\delta_{B)}^{2} 
\ , \ \ \ l^{\dot{A}\dot{B}} = 0
\end{equation}
The key function and the curvature
\begin{eqnarray}
 && \Theta = p^{\dot{2}} \, S(q^{\dot{2}}, q^{\dot{1}} p^{\dot{2}} p^{\dot{2}}) 
-\frac{1}{12} \frac{p^{\dot{2}}}{q^{\dot{1}}} (F_{0} \, p^{\dot{2}} - p^{\dot{1}})(F_{0} \, p^{\dot{2}} - 3p^{\dot{1}})  + \frac{1}{2} N \, p^{\dot{2}}  p^{\dot{2}}
\\ \nonumber
 &&     
C^{(2)} = F_{0} \, \frac{1}{q^{\dot{1}}q^{\dot{1}}} \ , \ \ C^{(1)} = -2 \, \frac{\partial}{\partial q^{\dot{1}}} \bigg( \frac{N}{2 q^{\dot{1}}}  - \frac{\partial N}{\partial q^{\dot{1}}} \bigg) - \frac{2 F_{0}  p^{\dot{1}}}{(q^{\dot{1}})^{3}} -   \frac{F_{0}^{2} p^{\dot{2}}}{(q^{\dot{1}})^{3}} \ \ \ \ \ \ 
\\ 
&& C_{\dot{A}\dot{B}\dot{C}\dot{D}} = \frac{1}{2} \dot{C}^{(1)} \delta^{\dot{2}}_{\dot{A}}\delta^{\dot{2}}_{\dot{B}}\delta^{\dot{2}}_{\dot{C}}\delta^{\dot{2}}_{\dot{D}}
\ , \ \  \dot{C}^{(1)} = 2 \, \Big( p^{\dot{2}} \, S(q^{\dot{2}}, q^{\dot{1}} p^{\dot{2}} p^{\dot{2}}) \Big)_{p^{\dot{2}}p^{\dot{2}}p^{\dot{2}}p^{\dot{2}}}
\end{eqnarray}
where $N=N(q^{\dot{M}})$ and $S=S(q^{\dot{2}}, q^{\dot{1}} p^{\dot{2}} p^{\dot{2}})$ are arbitrary functions, and $F_{0}$ is constant. The metric
\begin{eqnarray}
\label{metryka_typu_IIINxNlubnic}
ds^{2} &=& -2 \,  dp^{\dot{A}} \underset{s}{\otimes} dq_{\dot{A}}  
           -2 \, \bigg( (p^{\dot{2}} \, S)_{p^{\dot{2}}p^{\dot{2}}}- \frac{F_{0}^{2}}{2} \, \frac{p^{\dot{2}}}{q^{\dot{1}}} + N  \bigg) \, dq_{\dot{1}} \underset{s}{\otimes} dq_{\dot{1}}
\\ \nonumber
           &&+ \frac{p^{\dot{2}}}{ q^{\dot{1}}} \, dq_{\dot{2}} \underset{s}{\otimes} dq_{\dot{2}}
           +2 \bigg( 2F_{0} \, \frac{p^{\dot{2}}}{q^{\dot{1}}} -  \frac{p^{\dot{1}}}{q^{\dot{1}}} \bigg) \, dq_{\dot{1}} \underset{s}{\otimes} dq_{\dot{2}}
\end{eqnarray}
The cosmological constant must be necesarilly zero here $\Lambda=0$, hyperheavenly metrics are characterized by
\begin{itemize}
\item type $[\textrm{III}]^{n} \otimes [\textrm{N}]^{e}$: $F_{0} \ne 0$, $F_{0}$ can be re-gauged to $1$ without any loss of generality, $( p^{\dot{2}} \, S)_{p^{\dot{2}}p^{\dot{2}}p^{\dot{2}}p^{\dot{2}}} \ne 0$
\item type $[\textrm{N}]^{n} \otimes [\textrm{N}]^{e}$: $F_{0} = 0$, $C^{(1)} \ne 0$, $( p^{\dot{2}} \, S)_{p^{\dot{2}}p^{\dot{2}}p^{\dot{2}}p^{\dot{2}}} \ne 0$
\end{itemize}

The heavenly reductions of the metric obtained above are especially interesting here, because they offer two different null string geometries. One can get the metrics of the type $[\textrm{III,N}]^{n} \otimes [-]^{e}$ in which types $[\textrm{III,N}]$ corresponds to the nonexpanding self-dual null string and conformally flat, anti-self-dual part correspond to the expanding anti-self-dual null string. The second possible heavenly reduction $[-]^{n} \otimes [\textrm{N}]^{e}$ has different geometry of null strings. In this case expanding null string (anti-self-dual one) corresponds to the type $[\textrm{N}]$ and nonexpanding null string (self-dual one) corresponds to the conformally flat part. 

In order to obtain the heavenly metrics of the types $[\textrm{III,N}]^{n} \otimes [-]^{e}$, one must set $S(q^{\dot{2}}, q^{\dot{1}} p^{\dot{2}} p^{\dot{2}}) = f(q^{\dot{2}}) \, q^{\dot{1}} p^{\dot{2}}p^{\dot{2}}$ where $f$ is an arbitrary function of the variable $q^{\dot{2}}$. (It seems, that this is to strong condition and it is enough to set $( p^{\dot{2}} \, S)_{p^{\dot{2}}p^{\dot{2}}p^{\dot{2}}p^{\dot{2}}}=0$, but there is unused gauge freedom, which allows to simplify the function $S$). Heavenly metrics of the type $[-]^{n} \otimes [\textrm{N}]^{e}$ can be obtained by setting $F_{0}=0=N$ (once again condition $N=0$ is stronger then necessary $C^{(1)}=0$, but $N=0$ can be obtained by using gauge freedom). Finally
\begin{itemize}
\item type $[\textrm{III}]^{n} \otimes [-]^{e}$: $F_{0} \ne 0$, $F_{0}$ can be re-gauged to $1$ without any loss of generality, $S= f(q^{\dot{2}}) \, q^{\dot{1}} p^{\dot{2}}p^{\dot{2}}$ 
\item type $[\textrm{N}]^{n} \otimes [-]^{e}$: $F_{0} = 0$, $C^{(1)} \ne 0$, $S= f(q^{\dot{2}}) \, q^{\dot{1}} p^{\dot{2}}p^{\dot{2}}$ 
\item type $[-]^{n} \otimes [\textrm{N}]^{e}$: $F_{0}=0$, $N=0$, $( p^{\dot{2}} \, S)_{p^{\dot{2}}p^{\dot{2}}p^{\dot{2}}p^{\dot{2}}} \ne 0$
\end{itemize}
\textbf{Heavenly space of the type $[-]^{n} \otimes [\textrm{N}]^{e}$ - different approach}

Because the spaces $[-]^{n} \otimes [\textrm{N}]^{e}$ and $[\textrm{N}]^{e} \otimes [-]^{n}$ have the same null string geometry, there is another way to describe the heavenly metric, in which type $[\textrm{N}]$ is connected with  expanding null string and type $[-]$ is connected with nonexpanding null string. One can use the general theory of expanding hyperheavenly spaces and set $C_{\dot{A}\dot{B}\dot{C}\dot{D}}=0=C^{(3)}=C^{(2)}$. Such approach gives the Killing vector
\begin{equation}
K=\frac{\partial}{\partial \eta}
\end{equation}
and
\begin{equation}
l_{AB}= 0 \ \ \ , \ \ \ l^{\dot{A}\dot{B}} = - \frac{2}{\tau} \phi^{-1} J^{\dot{A}} J^{\dot{B}}
\end{equation}
The key function and the curvature
\begin{eqnarray}
 && W=\frac{1}{2} f(w,t) \, \phi^{2} 
\\ 
 &&  C^{(1)} = -2 \tau^{2} \phi^{7} f_{tt}   
\end{eqnarray}
where $f=f(w,t)$ is an arbitrary function and the metric reads
\begin{equation}
\label{metryka_typu_Nxnic_podejscieinne}
ds^{2} = \phi^{-2} \bigg\{ \frac{2}{\tau} (d \eta \underset{s}{\otimes} d w - d \phi \underset{s}{\otimes} dt) + 2  f \phi \, dw \underset{s}{\otimes}dw \bigg\}
\end{equation}
Of course, the metric (\ref{metryka_typu_Nxnic_podejscieinne}) is equivalent to the metric (\ref{metryka_typu_IIINxNlubnic}) with $F_{0}=0=N$.

\subsection{Spaces of the type $[\textrm{N},-]^{n} \otimes [\textrm{N},-]^{n}$}
\label{subsection_nullisometric_hyperheavenly_NxN_nn}

The last case is characterized by both self-dual and anti-self-dual null strings being nonexpanding. The only possible types are $[\textrm{N},-]^{n} \otimes [\textrm{N},-]^{n}$. These metrics have been found in \cite{biblio_dod_3}. The Killing vector takes the form
\begin{equation}
K= \frac{\partial}{\partial p^{\dot{1}}}
\end{equation}
with spinors $l_{AB}$ and $l_{\dot{A}\dot{B}}$
\begin{equation} 
l_{AB}=  0 \ \ \ , \ \ \ l^{\dot{A}\dot{B}} = 0 
\end{equation}
The key function and the curvature
\begin{eqnarray}
 && \Theta = \frac{1}{2} N(q^{\dot{M}}) \, p^{\dot{2}} p^{\dot{2}} + A(p^{\dot{2}}, q^{\dot{2}})
\\ 
    && 
 C^{(1)} = 2 \, \frac{\partial^{2} N}{\partial q^{\dot{1}} \partial q^{\dot{1}}} \ , \ \ \ 
\dot{C}^{(1)} = 2 \, A_{p^{\dot{2}}p^{\dot{2}}p^{\dot{2}}p^{\dot{2}}}
\end{eqnarray}
where $N=N(q^{\dot{M}})$ and $A=A(p^{\dot{2}}, q^{\dot{2}})$ are arbitrary functions. The metric reads
\begin{equation}
\label{metryka_typu_NxN_lubnic}
ds^{2} = -2 \,  dp^{\dot{A}} \underset{s}{\otimes} dq_{\dot{A}}  
           -2 \, \big( A_{p^{\dot{2}}p^{\dot{2}}}+ N  \big) \, dq_{\dot{1}} \underset{s}{\otimes} dq_{\dot{1}}
\end{equation}
A possible heavenly degenerations can be done in two different ways but they lead to the equivalent haevens. To get the heavenly space of the type $[-]^{n} \otimes [\textrm{N}]^{n}$ it is enought to set $N_{q^{\dot{1}}q^{\dot{1}}}=0$, then by using gauge freedom one can gauge $N$ away. The heavenly space of the type $[\textrm{N}]^{n} \otimes [-]^{n}$ can be obtained by setting the function $A$ as a third-order polynomial in $p^{\dot{2}}$ (in fact, taking into considerations the remaining gauge freedom, it is enough to set $A=f(q^{\dot{2}}) \, p^{\dot{2}}p^{\dot{2}}p^{\dot{2}}$). Gathering, we arrive at the cases
\begin{itemize}
\item type $[\textrm{N}]^{n} \otimes [\textrm{N}]^{n}$: $N_{q^{\dot{1}}q^{\dot{1}}} \ne 0$, $A_{p^{\dot{2}}p^{\dot{2}}p^{\dot{2}}p^{\dot{2}}} \ne 0$
\item type $[-]^{n} \otimes [\textrm{N}]^{n}$: $N = 0$, $A_{p^{\dot{2}}p^{\dot{2}}p^{\dot{2}}p^{\dot{2}}} \ne 0$
\item type $[\textrm{N}]^{n} \otimes [-]^{n}$: $N_{q^{\dot{1}}q^{\dot{1}}} \ne 0$, $A=f(q^{\dot{2}}) \, p^{\dot{2}}p^{\dot{2}}p^{\dot{2}}$
\end{itemize}

\section{Real slices.}

\subsection{Real slices with neutral signature $(++--)$}

All metrics presented in sections 4 and 5 are holomorphic. It is an easy matter to carry over all the results to the case of real spaces of the signature $(++--)$. Instead of the holomorphic objects (spinors, null strings, tetrads, coordinates, etc.) we deal with the real smooth objects. A real spaces with the neutral signature play an important role in Walker and Osserman geometry. A few papers dealing with such geometries appeared recently \cite{biblio_dod_17} - \cite{biblio_dod_12}. However, it was hyperheavenly formalism which allowed to obtain transparent results in Walker and Osserman geometries. For example, a new class of metrics admitting self-dual and anti-self-dual, parallely propagated null strings (\textsl{two-sided Walker spaces}) has been found in \cite{biblio_9}. These spaces have a natural generalization: if only one of the null strings is parallely propagated, we deal with so called \textsl{sesqui-Walker spaces}. Such spaces have been defined and investigated in \cite{biblio_dod_12}. Probably the most distinguished success of the hyperheavenly methods in Osserman geometry was finding all algebraically degenerate metrics of the globally Osserman space, which do not have the Walker property, i.e. they do not admit any parallely propagated null strings \cite{biblio_10}.

Some of the metrics presented in sections 4 and 5 are examples of the Walker or Osserman spaces admitting the null Killing vector. These metrics are gathered in the table:
\renewcommand{\arraystretch}{1.4}
\begin{displaymath}
\begin{tabular}{|l|l|}   \hline
sesqui-Walker & (\ref{metryka_homote_typpu_NxIII}), (\ref{metryka_homote_typpu_Nxnic}), (\ref{metryka_typu_IIINxNlubnic}) of the types $[\textrm{III,N,}-]^{n} \otimes [\textrm{N}]^{e}$ 
\\ \hline
two-sided Walker & (\ref{metryka_homote_typpu_IIIxnic}), (\ref{metryka_typu_IIINxNlubnic}) of the types $[\textrm{III,N,}-]^{n} \otimes [-]^{e}$, (\ref{metryka_typu_NxN_lubnic})
\\ \hline
globally Osserman & (\ref{metryka_homote_typpu_Nxnic}), (\ref{metryka_homote_typpu_IIIxnic}), (\ref{metryka_typu_IIIxIII}) of the type $[\textrm{N}]^{e} \otimes [-]^{e}$ or $[-]^{e} \otimes [\textrm{N}]^{e}$, 
\\ 
& (\ref{metryka_typu_IIINxNlubnic}) of the type $[\textrm{III,N,}-]^{n} \otimes [-]^{e}$ and $[-]^{n} \otimes [\textrm{N,}-]^{e}$, 
\\
& (\ref{metryka_typu_NxN_lubnic}) of the type $[\textrm{N}]^{n} \otimes [-]^{n}$ or $[-]^{n} \otimes [\textrm{N}]^{n}$
\\ \hline
\end{tabular}
\end{displaymath}
Among globally Osserman spaces the most interesting is that with metric (\ref{metryka_typu_IIIxIII}) of the type $[\textrm{N}]^{e} \otimes [-]^{e}$ or $[-]^{e} \otimes [\textrm{N}]^{e}$. It does not have any parallely propagated null strings (because of $\Lambda \ne 0$). Consequently it is the most general globally Osserman but not Walker space admitting a with null isometric Killing vector.

\subsection{Real slices with Lorentzian signature $(+++-)$}

Of course, the most interesting from the physical point of view are the Lorentzian slices with the signature $(+++-)$. There are still no general techniques of obtaining the Lorentzian slices, except some notes of properties of such slices \cite{biblio_5}. However, in some special cases such slices can be obtained quite easy.

It is well known, that there are only two subcases of the Einstein spaces with lorentzian signature and null Killing vector (compare \cite{biblio_dod_13}). One of them is pp-wave solution. The real metric of the pp-wave solution can be obtained from the complex metric (\ref{metryka_typu_NxN_lubnic}) of the type $[\textrm{N}]^{n} \otimes [\textrm{N}]^{n}$. Detailed discussion of this case can be found in \cite{biblio_dod_3}. We only mention, that it is enough to consider the necessary condition of existing real lorentzian slices, namely $C_{\dot{A}\dot{B}\dot{C}\dot{D}} = \bar{C}_{ABCD}$, or in this particular case: $\dot{C}^{(1)} = \bar{C}^{(1)}$.

Except the pp-wave solution, null Killing vector is admitted by the lorentzian, Einstein spaces of the type $[\textrm{II}]$ and $[\textrm{D}]$. It means, that desired lorentzian slice is hidden in hyperheavenly metric (\ref{metrykaprzestrzeni_IIxDD_oggolna}) with the key function (\ref{funkcja_kluczowa_dla_IIxII}) and with curvature given by (\ref{krzywizna_dla_hyperheavenly_isometricnull_IIxII}). Unfortunatelly, in this case the conditions $C_{\dot{A}\dot{B}\dot{C}\dot{D}} = \bar{C}_{ABCD}$ are not straightforward and technique which succeeded in pp-wave case, failed. 

However, one can consider the (complex) transformation (\ref{transformacja_lorentzowska_typu_IIxII}) which brings the metric to the form (\ref{metryka_typu_IIxII_wpostaci_lorr}). The metric (\ref{metryka_typu_IIxII_wpostaci_lorr}) depends on one (complex) function $M=M(x,y,u)$ which satisfies the equation (\ref{rownannie_typu_IIxII_wpostaci_lorr}). Treating now the coordinates $(x,y,u,v)$ as a real coordinates and the function $M$ as a real smooth function we find, that the metric (\ref{metryka_typu_IIxII_wpostaci_lorr}) automatically becomes real and has the Lorentzian signature. The vacuum Einstein equations have been reduced to the equation (\ref{rownannie_typu_IIxII_wpostaci_lorr}). Exactly the same form of the metric with the same equation describing real, vacuum, Lorentzian types $[\textrm{II}]$ and $[\textrm{D}]$ admitting null isometric Killing vector can be found in \cite{biblio_dod_13}. Summing up, the metric (\ref{metryka_typu_IIxII_wpostaci_lorr}) with real coordinates is another example of Lorentzian slice of the complex space. 

Why does this technique succeed? The first reason is, probably, the explicit using of the imaginary unit in transformation (\ref{transformacja_lorentzowska_typu_IIxII}). It plays no role if we consider (\ref{transformacja_lorentzowska_typu_IIxII}) as complex transformation and the coordinates $(x,y,u,v)$ as complex. But if $(x,y,u,v)$ are real, this step changes automatically the signature of the metric making (\ref{metryka_typu_IIxII_wpostaci_lorr}) Lorentzian. The second reason is that the metric (\ref{metryka_typu_IIxII_wpostaci_lorr}) does not depend directly on the key function $W$, but on the function $M$. The relation between this two functions is given by (\ref{definicja_function_M}) and it contains the imaginary unit. However, by twice differentiation the hyperheavenly equation (\ref{hiperniebianskie_dla_typu_IIxII_formalizm_XY}) can be brought to the form (\ref{rownannie_typu_IIxII_wpostaci_lorr}) which is free of imaginary unit. Finaly, we are left with the real metric and the real equation. It is enough to finish the contruction of the Lorentzian spaces of the type $[\textrm{II}]$ and $[\textrm{D}]$ admitting null Killing vector.

It is worth to note, that this construction succeeds only in the vacuum case. If cosmological constant $\Lambda \ne 0$ we have not been able to find the Lorentzian slice. 

No spaces with null homothetic symmetries admit Lorentzian slices.

\section{Concluding remarks.}

In the presented paper the null Killing vectors (isometric and homothetic) in complex spacetime have been considered. The connection between null Killing vectors and null strings has been pointed in section 2. Because of the existence of the null strings the most natural apparatus in investigating null Killing vectors appeared to be hyperheavenly and heavenly spaces. After short summary of the structure of hyperheavenly spaces (section 3), we have been able to present all possible metrics admitting null Killing vector. Only two of them
\begin{itemize}
\item the metric (\ref{metryka_homote_typpu_NxIII}) of the type $[\textrm{N}]^{e} \otimes [\textrm{III}]^{n}$ with null homothetic Killing vector
\item the metric (\ref{metrykaprzestrzeni_IIxDD_oggolna}) of the types $[\textrm{II}]^{e} \otimes [\textrm{II}]^{e}$ with $\Lambda \ne 0$ and with null isometric Killing vector
\end{itemize}
have not been solved completely. In (\ref{metryka_homote_typpu_NxIII}) the functions $F=F(x,w,t)$ and $\gamma=\gamma(w,t)$ satisfy the equation (\ref{hyperheavenlyeq_nullhomothetic}). No solution with $F_{xxx} \ne 0$ and $\gamma_{t} \ne 0$ have been found. However, the geometry of this space is so interesting and the type $[\textrm{N}]^{e} \otimes [\textrm{III}]^{n}$ so rare, that we are going to study the equation (\ref{hyperheavenlyeq_nullhomothetic}) with detailes in future. What concerns the metric (\ref{metrykaprzestrzeni_IIxDD_oggolna}) we have been able to solve the case $[\textrm{D}]^{e} \otimes [\textrm{D}]^{e}$ and reduce the type $[\textrm{II}]^{e} \otimes [\textrm{II}]^{e}$ with $\Lambda=0$ to the Euler-Poisson-Darboux equation which solutions are known. The type $[\textrm{II}] \otimes [\textrm{II}]$ with $\Lambda \ne 0$ has been reduced to the equation (\ref{typ_IIxII_hyperheavenly_equation_red3}), but this reduction has obvious disadvantages. Like in the previous case, we will deal with this equation soon.

The transparent results are the metrics (\ref{metryka_homote_typpu_Nxnic}) and (\ref{metryka_homote_typpu_IIIxnic}) which constitute all heavens with null homothetic symmetry. These cases have been considered in \cite{biblio_dod_14}, but without giving any explicit form of the metric. We were able to integrate the problem completely.

Probably the most interesting from the physical point of view is another example of real Lorentzian slice of the complex metric. The first such an example has been presented in \cite{biblio_dod_3}. Here we have been able to find the Lorentzian slices of the types $[\textrm{II}]^{e} \otimes [\textrm{II}]^{e}$ and $[\textrm{D}]^{e} \otimes [\textrm{D}]^{e}$ with $\Lambda=0$. They are given by the metric (\ref{metryka_typu_IIxII_wpostaci_lorr}) which depends on one function $M$ of three variables satisfying the equation (\ref{rownannie_typu_IIxII_wpostaci_lorr}). Such a metric has been presented earlier (see \cite{biblio_dod_15}, or in a concise form \cite{biblio_dod_13}). 

Both these examples gave some valuable notes about obtaining Lorentzian slices. In the first of them the condition $C_{\dot{A}\dot{B}\dot{C}\dot{D}} = \bar{C}_{ABCD}$ has been succesfully used, in the second one a resonable using of imaginary unit appears to be essential. Nonetheless, the case with nonzero cosmological constant is still unsolved. Taking into considerations the optical properties of the congruneces of null geodesics defined by the null isometric Killing vector (\ref{geodesic_expansion}) - (\ref{geodesic_shear}) we conclude that all such slices must belong to the Kundt class (\cite{biblio_dod_13}, xxxi). The vacuum Einstein field equations with cosmological constant for the Kundt class have been gathered in \cite{biblio_dod_13} but no explicit solution has been presented there. The vacuum types [II] and [D] with cosmological constant admitting null isometric Killing vector via Lorentzian slices of complex types $[\textrm{II}]^{e} \otimes [\textrm{II}]^{e}$ and $[\textrm{D}]^{e} \otimes [\textrm{D}]^{e}$ become our main issue to be considered in future. 

However, interesting observation can be immediately made. The generic complex spacetimes admitting null isometric Killing vector which have the Lorentzian slices are equipped with null strings which have the same properties. Indeed, pp-wave solution can be obtained from the complex spacetime of the type $[\textrm{N}]^{n} \otimes [\textrm{N}]^{n}$ where both self-dual and anti-self-dual congruences of null strings are nonexpanding. In the opposite, Lorentzian types [II] and [D] have been obtained from the complex spacetimes of the types $[\textrm{II}]^{e} \otimes [\textrm{II}]^{e}$ and $[\textrm{D}]^{e} \otimes [\textrm{D}]^{e}$ in which both congruences of null strings are expanding. Maybe it is a general rule and Lorentzian slices can be obtained only from the complex spacetimes equipped with both expanding or both nonexpanding null strings? 

We hope, that further investigation of the structure of complex spacetime allows to find some effective and more general techniques of obtaining real Lorentzian slices.


\begin{thebibliography}{99}





\bibitem{biblio_1} Plebański J F and Robinson I 1976 Left - degenerate vacuum metrics \textsl{Phys. Rev. Lett.} \textbf{37} 493



\bibitem{biblio_4} Finley III J D and Plebański J F 1976 The intrinsic spinorial structure of hyperheavens \textsl{J. Math. Phys.} \textbf{17} 2207

\bibitem{biblio_2} Plebański J F and Robinson I 1977 The complex vacuum metric with minimally degenerated conformal curvature \textsl{Asymptotic Structure of Space-Time} eds. by Esposito F P and Witten L 
(Plenum Publishing Corporation, New York, 1977) pp. 361-406.


\bibitem{biblio_3} Boyer C P, Finley III J D and Plebański J F 1980 Complex general relativity, $\mathcal{H}$ and $\mathcal{HH}$ spaces - a survey to one approach in: \textsl{General Relativity and Gravitation. Einstein Memorial Volume} ed. A. Held (Plenum, New York) vol.2. pp. 241-281



\bibitem{biblio_17} Plebański J F and Hacyan S 1976 Some properties of Killing spinors \textsl{J. Math. Phys.} \textbf{14} 2204


\bibitem{biblio_8} Finley J D III and Plebański J F 1979 The clasification of all $\mathcal{H}$ spaces admitting a Killing vector \textsl{J. Math. Phys.} \textbf{20} 1938


\bibitem{biblio_7} Plebański J F and Finley J D 1978 Killing vectors in nonexpanding HH spaces \textsl{J. Math. Phys.} \textbf{19} 760


\bibitem{biblio_dod_2} Sonnleitner S A and Finley J D III 1982 The form of Killing vectors in expanding $\mathcal{HH}$ spaces \textsl{J. Math. Phys.} \textbf{23(1)} 116




\bibitem{biblio_dod_16} Garc\'\i a D A and Plebański J F 1977 Seven parametric type - D solutions of Einstein - Maxwell equations in the basic left - degenerate representation \textsl{Il Nuovo Cimento} Serie 11 \textbf{40} B 224




\bibitem{biblio_dod_30} Robinson D C 1987 Some Real and Complex Solutions of Einstein's Equations \textsl{General Relativity and Gravitation} \textbf{19} 7


\bibitem{biblio_dod_31} Hickman M S and McIntosh C B 1986 Complex relativity and real solutions. III. Real type-N solutions from complex N$\otimes$N ones \textsl{General Relativity and Gravitation} \textbf{18} 107

\bibitem{biblio_dod_32} Plebański J F, Przanowski M and Formański S 1998 Linear superposition of two type-N nonlinear gravitons \textsl{Physics Letters A} \textbf{246} 25


\bibitem{biblio_dod_33} Robinson D C 2002 Holomorphic 4-Metrics and Lorentzian Structures \textsl{General Relativity and Gravitation} \textbf{34} No 8 1173





\bibitem{biblio_5} Rózga K 1977 Real slices of complex space - time in general relativity \textsl{Rep. Math. Phys.} \textbf{11} 197



\bibitem{biblio_9} Chudecki A and Przanowski M 2008 From hyperheavenly spaces to Walker and Osserman spaces: I \textsl{Class. Quantum Grav.} \textbf{25} 145010



\bibitem{biblio_10} Chudecki A and Przanowski M 2008 From hyperheavenly spaces to Walker and Osserman spaces: II \textsl{Class. Quantum Grav.} \textbf{25} 235019



\bibitem{biblio_15} Chudecki A 2010 Conformal Killing vectors in Nonexpanding $\mathcal{HH}$-spaces with $\Lambda$ \textsl{Class. Quantum Grav.} \textbf{27} 205004


\bibitem{biblio_dod_3} Chudecki A 2012 Classification of the Killing vectors in nonexpanding $\mathcal{HH}$-spaces with $\Lambda$ \textsl{Class. Quantum Grav.} \textbf{29} 135010

\bibitem{biblio_16} Chudecki A 2013 Homothetic Killing Vectors in Expanding $\mathcal{HH}$-spaces with $\Lambda$ \textsl{International Journal of Geometric Methods in Modern Physics} Vol 10 No 1 1250077


\bibitem{biblio_dod_10} Chudecki A and Przanowski M 2013 Killing symmetries in $\mathcal{H}$-spaces with $\Lambda$ to be published











\bibitem{biblio_dod_11} Koshlyakov N S, Smirnov M M and Gliner E B 1964 Differential Equations of Mathematical Physics \textsl{North Holland Publishing Company} Amsterdam





\bibitem{biblio_dod_17} Law P R and Matsushita Y 2008 A spinor approach to Walker geometry \textsl{Communications in Mathematical Physics} \textbf{282} (3) 577-623


\bibitem{biblio_dod_18} Law P R and Matsushita Y 2011 Algebraically Special Real Alpha Geometries \textsl{Journal of Geometry and Physics} \textbf{61} 2064-2080


\bibitem{biblio_dod_12} Law P R and Matsushita Y 2013 Real AlphaBeta-geometries and Walker geometry \textsl{Journal of Geometry and Physics} \textbf{65} 35-44






\bibitem{biblio_dod_13} Stephani H, Kramer D, MacCallum M A H, Hoenselaers C and Herlt E 2003 Exact Solutions to Einstein's Field Equations. Second Edition \textsl{Cambridge University Press, Cambridge}



\bibitem{biblio_dod_14} Dunajski M and West S. 2007 Anti-Self-Dual Conformal Structures with Null Killing Vectors from Projective Structures , Commun. Math. Phys. 272, 85-118


\bibitem{biblio_dod_15} Dautcourt G 1964 Gravitationsfelder mit isotropem Killingvektor \textsl{Relativistic
theories of gravitation} ed. L. Infeld, page 300 (Pergamon Press, Oxford).










\end{thebibliography}
\end{document}